\documentclass[twocolumn,superscriptaddress,prb,preprintnumbers,nobibnotes,aps,showkeys]{revtex4-2}  
\usepackage{graphicx}
\usepackage[caption=false,font=large]{subfig}
\usepackage{floatrow}
\usepackage{epstopdf}
\usepackage{tabularx}
\usepackage{dcolumn}
\usepackage{multirow}
\usepackage{hyperref}
\hypersetup{
  colorlinks  = true, 
  urlcolor   = blue, 
  linkcolor  = blue, 
  citecolor  = blue 
}
\usepackage{xcolor}
\usepackage{soul}
\usepackage{natbib}
\usepackage{amsmath}
\usepackage{rotating}
\setcitestyle{square,numbers}

\newcommand{\gc}{$^\circ$C}

\begin{document}

\title{Crystal Growth, Exfoliation and Magnetic Properties of Quaternary Quasi-Two-Dimensional CuCrP$_2$S$_6$}	

\author{S.~Selter}
\affiliation{Institute for Solid State Research, Leibniz IFW Dresden, Helmholtzstr. 20, 01069 Dresden, Germany}
\author{K.~K.~Bestha}
\affiliation{Institute for Solid State Research, Leibniz IFW Dresden, Helmholtzstr. 20, 01069 Dresden, Germany}
\author{P.~Bhattacharyya}
\affiliation{Institute for Theoretical Solid State Physics, Leibniz IFW Dresden, Helmholtzstr. 20, 01069 Dresden, Germany}
\author{B.~\"{O}zer}
\affiliation{Institute for Solid State Research, Leibniz IFW Dresden, Helmholtzstr. 20, 01069 Dresden, Germany}
\author{Y.~Shemerliuk}
\affiliation{Institute for Solid State Research, Leibniz IFW Dresden, Helmholtzstr. 20, 01069 Dresden, Germany}
\author{L.~T.~Corredor}
\affiliation{Institute for Solid State Research, Leibniz IFW Dresden, Helmholtzstr. 20, 01069 Dresden, Germany}
\author{L.~Veyrat}
\affiliation{Institute for Solid State Research, Leibniz IFW Dresden, Helmholtzstr. 20, 01069 Dresden, Germany}
\affiliation{Institute of Solid State and Materials Physics and W\"urzburg-Dresden Cluster of Excellence ct.qmat, Technische Universit\"at Dresden, 01062 Dresden, Germany}
\author{A.~U.~B.~Wolter}
\affiliation{Institute for Solid State Research, Leibniz IFW Dresden, Helmholtzstr. 20, 01069 Dresden, Germany}
\author{L.~Hozoi}
\affiliation{Institute for Theoretical Solid State Physics, Leibniz IFW Dresden, Helmholtzstr. 20, 01069 Dresden, Germany}
\author{B.~B\"{u}chner}
\affiliation{Institute for Solid State Research, Leibniz IFW Dresden, Helmholtzstr. 20, 01069 Dresden, Germany}
\affiliation{Institute of Solid State and Materials Physics and W\"urzburg-Dresden Cluster of Excellence ct.qmat, Technische Universit\"at Dresden, 01062 Dresden, Germany}
\author{S.~Aswartham}
\affiliation{Institute for Solid State Research, Leibniz IFW Dresden, Helmholtzstr. 20, 01069 Dresden, Germany}
\date{\today}

\begin{abstract}
We report optimized crystal growth conditions for the quaternary compound CuCrP$_2$S$_6$ by chemical vapor transport. Compositional and structural characterization of the obtained crystals were carried out by means of energy-dispersive X-ray spectroscopy and powder X-ray diffraction. CuCrP$_2$S$_6$ is structurally closely related to the $M_2$P$_2$S$_6$ family ($M$: transition metal), which contains several compounds that are under investigation as 2D magnets. As-grown crystals exhibit a plate-like, layered morphology as well as a hexagonal habitus. We present successful exfoliation of such as-grown crystals down to thicknesses of 2.8~nm corresponding to 4 layers. CuCrP$_2$S$_6$ crystallizes in the monoclinic space group $C2/c$. Magnetization measurements reveal an antiferromagnetic ground state with $T_\textrm{N} \approx 30$\,K and a positive Curie-Weiss temperature in agreement with dominant ferromagnetic intralayer coupling. Specific heat measurements confirm this magnetic phase transition and the magnetic order is suppressed in an external magnetic field of about 6\,T (8\,T) applied parallel (perpendicular) to the $ab$ plane. At higher temperatures between 140-200~K additional broad anomalies associated with structural changes accompanying antiferroelectric ordering are detected in our specific heat studies. 
\end{abstract}


\maketitle

\section{Introduction}

For future generations of functional devices in spintronics and quantum technology, two-dimensional (2D) materials are of great interest \cite{AGeim2013,CGong2019,Wang2018,Gibertini2019,Zhong2017,LCasto2015,Samarth2017}. With their potential for such applications in mind, the electronic properties of various 2D materials were extensively investigated following the experimental discovery of graphene. Materials such as the family of transition metal dichalcogenides (TMCs), which exhibit a wide range of band gaps in mono- and few-layer samples, were successfully used in optoelectronic~\cite{Baugher2014}, spintronic \cite{Li2019} and photovoltaic devices \cite{Fan2019}, among others.
However, for more advanced functionalities, 2D materials with intrinsic magnetic and ferroic properties are necessary, which the aforementioned materials lack. Consequently, the search for magnetic 2D materials and their detailed investigation have lately moved into the focus of 2D materials research \cite{CGong2017,BHuang2017,Song2019,Wang2018a}.

\begin{figure*}[htb]
\centering
\includegraphics[width=\columnwidth]{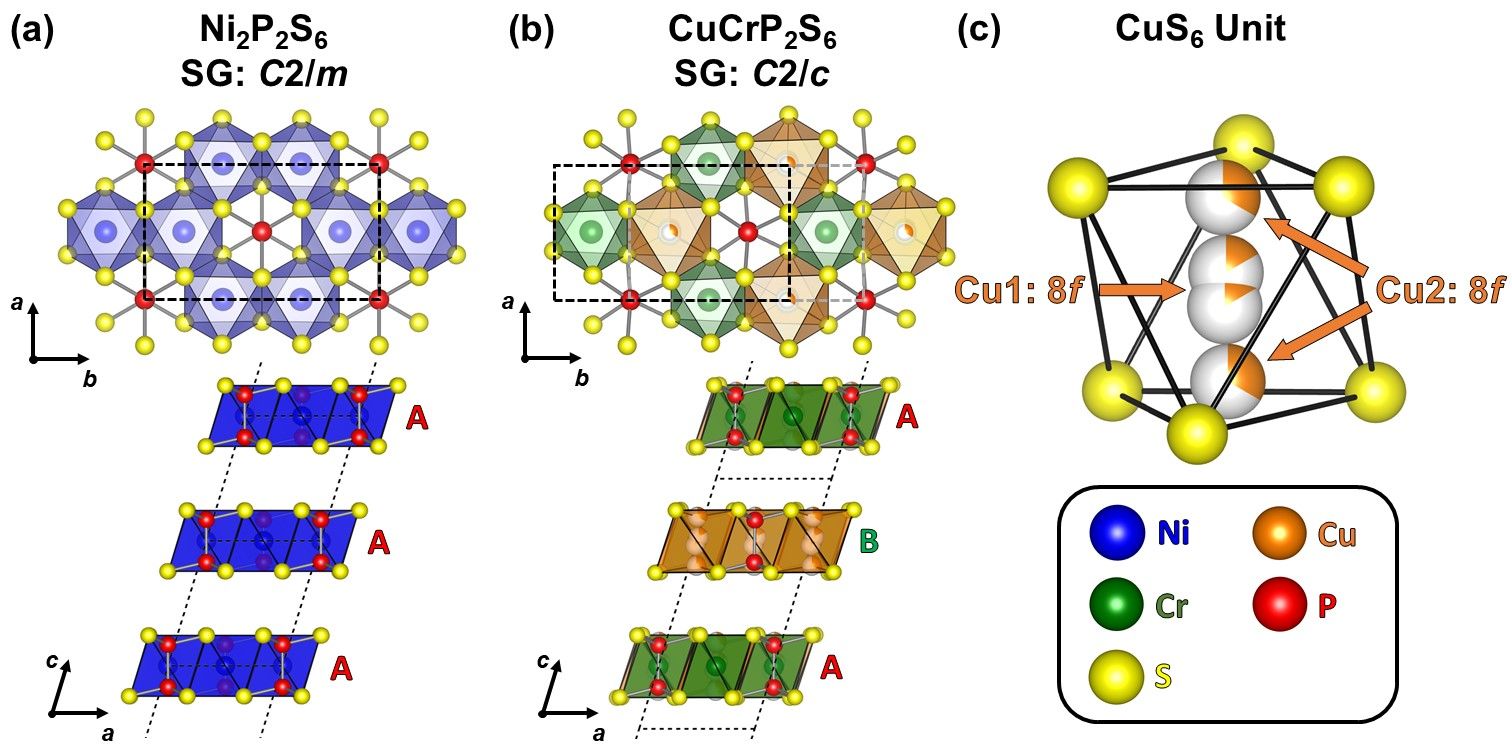}
\caption{
Comparison of the structures of (a) Ni$_2$P$_2$S$_6$ \cite{SSelter2020} (as exemplary member of the ternary $M_2$P$_2$S$_6$ compounds) and (b) CuCrP$_2$S$_6$. Top: view on the $ab$ plane; bottom: view along $b$ illustrating the monoclinic stacking of layers in the $ac$ plane. The dashed lines show the contour of the unit cell. The CuS$_6$ structural unit is shown separately in (c) to illustrate the different Cu positions in CuCrP$_2$S$_6$ according to the structural model proposed by Colombet \textit{et al.}~\cite{PColombet1982}, with the partial filling of the Cu spheres corresponding to their site occupancies.
}
\label{fig:Intro_Structure}
\end{figure*}

The material family of transition metal phosphorus sulfides with the general composition $M_2$P$_2$S$_6$ ($M$: transiton metal) is in this respect of big interest as it combines a van-der-Waals layered 2D crystal structure, as illustrated in Fig.~\ref{fig:Intro_Structure} (a), with magnetic properties depending on the transition element cation $M^{2+}$ \cite{Brec1985,Susner2017}. In this material class, the quaternary $M^{1+}M’^{3+}$P$_2$S$_6$ compounds, in which $M^{2+}$ is replaced by $M^{1+}_{0.5}M’^{3+}_{0.5}$, are especially interesting, as the two differently charged transition element cations order on the honeycomb sublattice. They either form an alternating or a stripe-like pattern depending on the difference in the ionic radius of $M^{1+}$ and $M’^{3+}$ \cite{Brec1985,PColombet1982,PColombet1983,Selter2021a}.
For example, the former arrangement is observed for AgInP$_2$S$_6$~\cite{Ouili1987} (ionic radii~\cite{RShannon1976}: Ag$^{1+}$: 1.15\,\r{A}; In$^{3+}$: 0.80\,\r{A}) and the latter for AgCrP$_2$S$_6$ \cite{PColombet1983,Selter2021a} (ionic radii: Ag$^{1+}$: 1.15\,\r{A}; Cr$^{3+}$: 0.62\,\r{A}).

The title compound CuCrP$_2$S$_6$ (ionic radii: Cu$^{1+}$: 0.77\,\r{A}; Cr$^{3+}$: 0.62\,\r{A}) also exhibits an alternating arrangement of Cu and Cr on the honeycomb sublattice, however, with an additional structural feature (Fig. ~\ref{fig:Intro_Structure} (b), (c)). In other $M_2$P$_2$S$_6$ and $M^{1+}M’^{3+}$P$_2$S$_6$ compounds, the transition element cations of one layer are located in one common plane, such that each cation is in an octahedral coordination environment of surrounding sulfur atoms. In CuCrP$_2$S$_6$, the Cr cations are also located at this position, but the Cu cations are strongly displaced either towards the bottom or top layer of sulfur and a nearly trigonal planar coordination environment. This displacement of the Cu position away from the center of their octahedral S$_6$ environment gives rise to a local electric dipole moment \cite{PColombet1982,Lai2019,Susner2020}.

At room temperature, the occupation of the top and bottom Cu positions is random resulting in an overall paraelectric behavior. However, below about 270\,K ordering of the Cu atoms sets in, which leads to a long-range antipolar order of Cu atoms below 140\,K and to an antiferroelectric phase. The complex nature of the antiferroelectric/paraelectric phase transition was recently investigated by Susner \textit{et al.} by temperature dependent Raman scattering and x-ray diffraction experiments \cite{Susner2020}. While the different electric phases are linked to the Cu sublattice, the Cr sublattice with Cr$^{3+}$ having S = 3/2 gives rise to an antiferromagnetic ground state with an ordering temperature of $T_\textrm{N} = 30$\,K~\cite{PColombet1982}. This makes CuCrP$_2$S$_6$ an exciting multi(anti-)ferroic compound in which the different ferroic contributions are linked to independent sublattices. In addition, theoretical investigations indicate that the application of an external electric field is sufficient to stabilize a ferroelectric phase instead of the antiferrolectric arrangement \cite{Lai2019}.

Although CuCrP$_2$S$_6$ has recently moved into the focus of materials research due to its interesting ferroelectric properties in combination with a low-dimensional structure, a detailed investigation and understanding of its magnetic properties is missing until now. Colombet \textit{et al.}, who were first to synthesize and investigate this material, already reported the observation of antiferromagnetic order at low temperatures in bulk crystals \cite{PColombet1982} in agreement with more recent reports \cite{WKleemann2011,Susner2020}. However, the coexistence of dominantly ferromagentic interactions and an antiferromagnetic groundstate that has been observed for CuCrP$_2$S$_6$ calls for further investigation in order to fully understand the magnetism and, further on, its implications on the ferroic properties in this  2D material.

Consequently, we report here the optimized growth of CuCrP$_2$S$_6$ bulk single crystals via chemical vapor transport, their detailed structural and compositional characterization as well as a thorough investigation of the magnetic properties based on magnetometry and specific heat measurements. Based on quantum chemical calculations, we extract the Cr$^{3+}$ $3d^3$ multiplet structure and investigate the influence of different Cu-ion configurations on the Cr-site crystal field splittings. Furthermore, we show first results from the mechanical exfoliation of bulk CuCrP$_2$S$_6$ crystals demonstrating that even layers with a thickness of 2.8\,nm can be produced and are stable in air, and which represents the first vital step towards their application in spintronics or quantum technology.

\section{Materials \& Methods}

The chemical educts for the crystal growth of CuCrP$_2$S$_6$ were obtained from Alfa Aesar and are listed in Table~\ref{tab:educts}. All chemical educts were kept under argon atmosphere in a glove box for storage and handling. The crystal growth was carried out in a custom-made two-zone quartz tube furnace under constant argon flow of approx. $10^{-2}$ l/h. Temperatures were monitored and adjusted by two separate temperature controllers from Nabertherm.

\begin{table}[htb]
    \centering
    \begin{ruledtabular}
    \begin{tabularx}{\columnwidth}{lld}
    \multicolumn{1}{c}{\textbf{Substance}} & \multicolumn{1}{c}{\textbf{Specification}} & \multicolumn{1}{c}{\textbf{Purity}} \\
    Copper & Powder, -100+325 mesh & 99.9\% \\
    Chromium & Powder, -100+325 mesh & 99.99\%\\
    Phosphorus & Lumps, red & 99.999\%\\
    Sulfur & Pieces & 99.999\%\\
    Iodine & Resublimed crystals & 99.9985\%\\
    \end{tabularx}
    \end{ruledtabular}
    \caption{Chemical educts for the crystal growth of CuCrP$_2$S$_6$ by the chemical vapor transport technique.}
    \label{tab:educts}
\end{table}

The crystals grown in this work were characterized by scanning electron microscopy (SEM) regarding their morphology and topography using a secondary electron (SE) detector and regarding chemical homogeneity via the chemical contrast obtained from a back scattered electron (BSE) detector. For this, a ZEISS EVO MA 10 scanning electron microscope was used. The chemical composition of the crystals was investigated by energy dispersive X-ray spectroscopy (EDX), which was measured in the same SEM device with an accelerating voltage of 30\,kV for the electron beam and using an energy dispersive X-ray analyzer.

The crystallographic phase and structure of the grown crystals was investigated by powder X-ray diffraction (pXRD) at room temperature. For this, a fine powder was prepared from the crystals in an agate mortar. A STOE STADI laboratory diffractometer in transmission geometry with Cu-K$_{\alpha1}$ radiation from a curved Ge(111) single crystal monochromator and a MYTHEN 1K 12.5$^\circ$-linear position sensitive detector manufactured by DECTRIS was used for the measurement of the pXRD patterns. These patterns were initially analyzed by pattern matching using the HighScore Plus program suite~\cite{Degen2014}. Once the crystallographic phase was identified by pattern matching, a refinement of the crystal structure model was performed based on the experimental patterns using the Rietveld method in Jana2006~\cite{Petricek2014}.

The details of the Cr$^{3+}$ 3$d^3$ multiplet structure in CuCrP$_2$S$_2$ are accessible with quantum chemical electronic-structure computational methods \cite{qc_book_00}. Embedded-cluster complete-active-space self-consistent-field (CASSCF) \cite{qc_book_00} and post-CASSCF configuration-interaction (CI) computations were performed in this regard using the quantum chemical package {\sc molpro} \cite{Molpro}.

Field and temperature dependent magnetometry was performed using a superconducting quantum interference device vibrating sample magnetometer (SQUID-VSM) by Quantum Design (MPMS3). Specific heat measurements were performed on a single crystal of mass m = 1.9(3) mg  using a heat-pulse relaxation method in a Physical Property Measurement System (PPMS) from Quantum Design. For temperature ranges 2-140 K and 135-220 K, Apiezon N and H grease were used, respectively.
In order to obtain the intrinsic specific heat of CuCrP$_2$S$_6$, the temperature- and field-dependent addenda were thoroughly subtracted from the measured specific-heat values in the sample measurements.

\section{Results}

\subsection{Crystal Growth by Chemical Vapor Transport}

\begin{sloppypar} The elemental educts copper, chromium, red phosphorus and sulfur were weighed out to yield a molar ratio of Cu\,:\,Cr\,:\,P\,:\,S~=~1\,:\,1\,:\,2\,:\,6 and homogenized in an agate mortar. 0.5\,g of the reaction mixture were loaded in a quartz ampule (6\,mm inner diameter, 2\,mm wall thickness) together with approx. 50\,mg of the transport agent iodine. Immediately prior to use, the ampule was cleaned with water, rinsed with isopropanol and, thereafter, baked out at 800\,$^\circ$C in an electric tube furnace for at least 12\,h. This is done to avoid contamination of the reaction volume with traces of water. The preparation of the reaction mixture as well as the filling of the ampule were carried out inside of an argon-filled glove box.\end{sloppypar}

\begin{figure}[htb]
    \centering
    \includegraphics[width=\columnwidth]{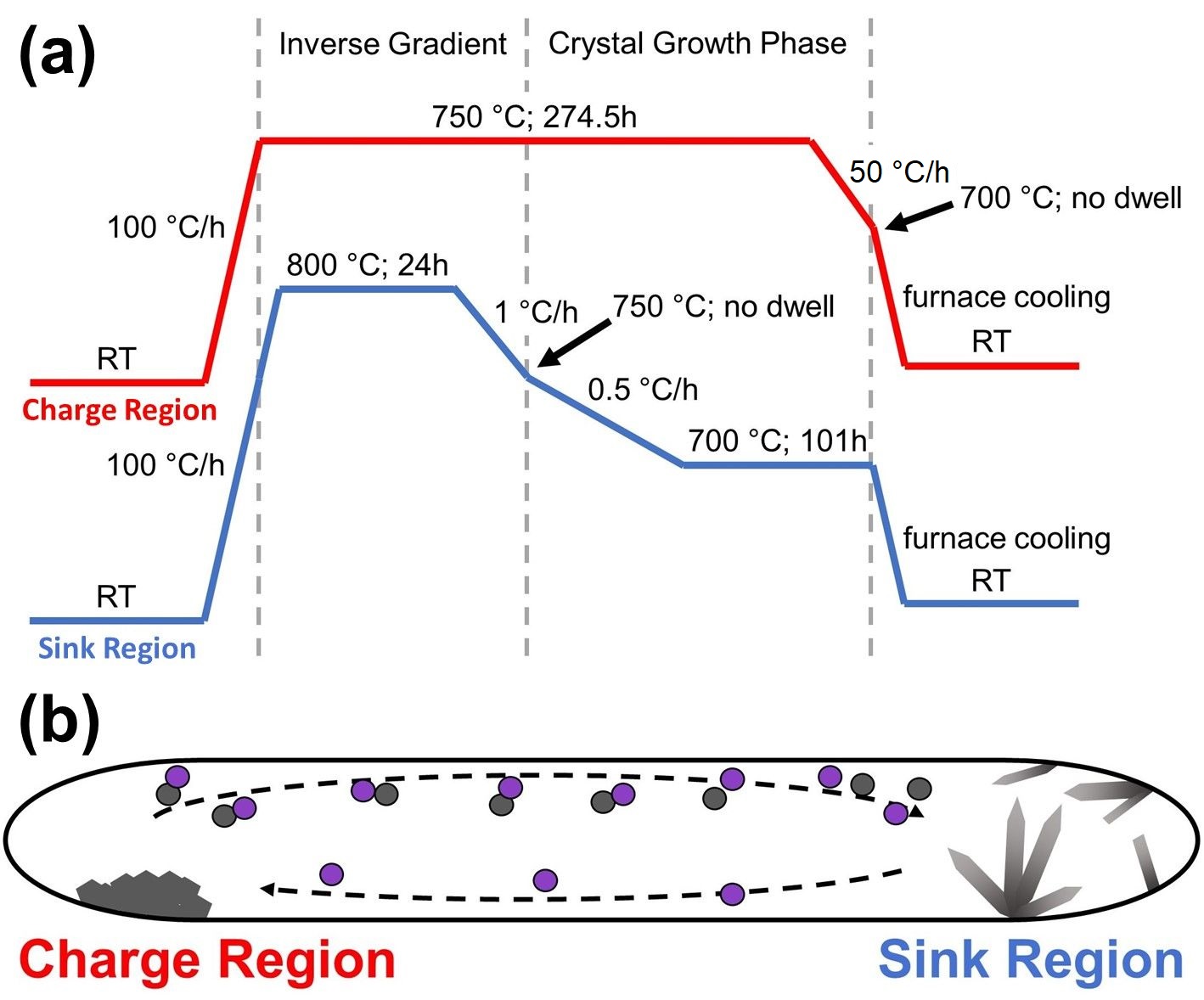}
    \caption{(a) Graphical illustration of the temperature profile for the CVT crystal growth of CuCrP$_2$S$_6$ and (b) schematic drawing of an ampule during CVT. Arrows indicate the mass flow of the volatile transport species (top) and the flow of the released transport agent back to the charge (bottom) \cite{Selter2021a}.}
    \label{fig:TProfile}
\end{figure}

The filled ampule was then transferred to a vacuum pump and evacuated to a residual pressure of 10$^{-8}$\,bar. To suppress the unintended sublimation of the transport agent during this step, the lower end of the ampule containing the material was cooled with a small Dewar flask filled with liquid nitrogen. After the desired internal pressure was stabilized, the valve to the vacuum pump was closed, the cooling was stopped and the ampule was sealed under static pressure at a length of approximately 12 cm.

The ampule was placed horizontally in a two-zone tube furnace in such a way that the elemental mixture was only at one side of the ampule which is called the charge side. As illustrated in Fig.~\ref{fig:TProfile}, the furnace was heated homogeneously to 750\,\gc\ at 100\,\gc/h. The charge side was kept at this temperature for 274.5\,h, while the other side of the ampule which is the sink side was initially heated up to 800\,\gc\ at 100\,\gc/h, dwelled at this temperature for 24\,h and then cooled back to 750\,\gc\ at 1\,\gc/h. An inverse transport gradient is formed, \textit{i.e.} transport from sink to charge, to clean the sink side of particles which stuck to the walls of the quartz ampule during filling. This ensures an improved nucleation condition in the following step. Then the sink side was cooled to 700\,\gc\ at 0.5\,\gc/h to slowly form the thermal transport gradient resulting in a controlled nucleation. As a next step, the ampule was dwelled with a transport gradient of 750\,\gc\ (charge) to 700\,\gc\ (sink) for 100\,h. After this the charge side was cooled to the sink temperature in 1\,h before both sides were furnace-cooled to room temperature. Similar conditions were successfully used to grow crystals of the sister compound AgCrP$_2$S$_6$ \cite{Selter2021a} and crystals of the ternary $M_2$P$_2$S$_6$ compounds \cite{Selter2021,Dioguardi2020}.

\begin{figure}[htb]
    \centering
    \includegraphics[width=\columnwidth]{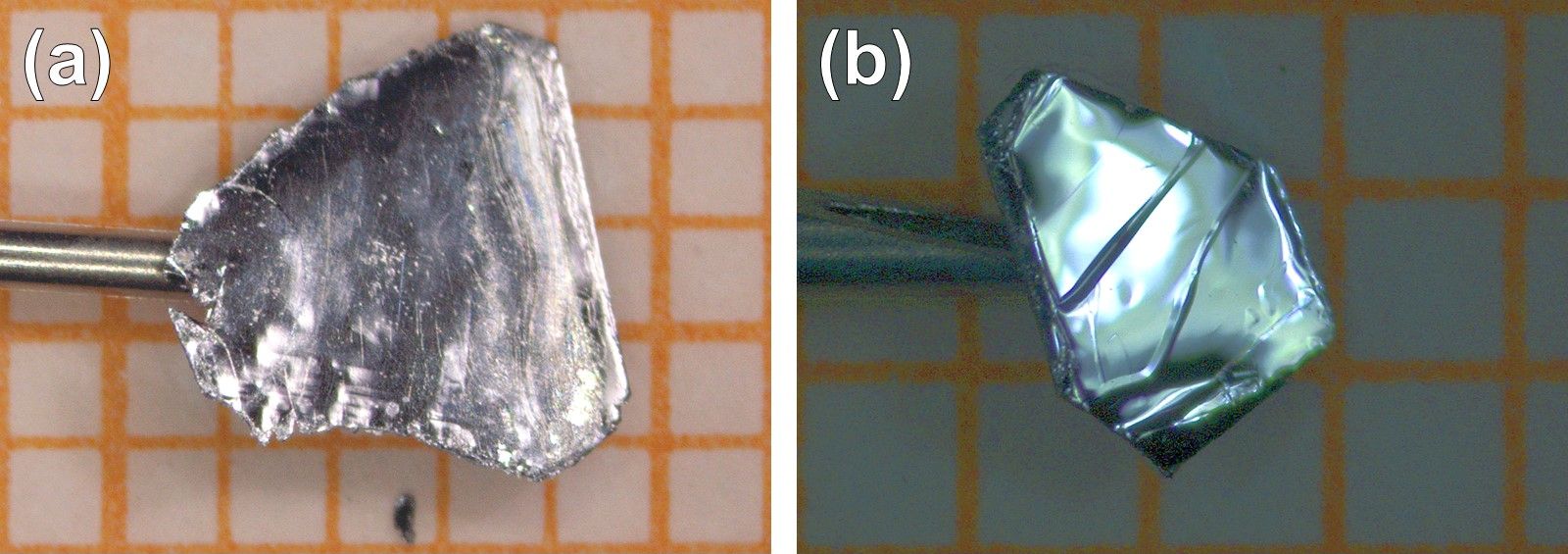}
    \caption{(a) As-grown and (b) freshly exfoliated crystals of CuCrP$_2$S$_6$. An orange square in the background is $1 \times 1$\,mm$^2$ for scale.}
    \label{fig:CrystalImages}
\end{figure}

Using this procedure, shiny plate-like crystals of CuCrP$_2$S$_6$ in the size of up to 3\,mm~$\times$~3\,mm~$\times$~200\,$\mu$m were obtained. The as-grown single crystals, as shown in Fig.~\ref{fig:CrystalImages}(a), exhibit a layered morphology and are easy to exfoliate. A freshly exfoliated crystal with a highly reflective surface is shown in Fig.~\ref{fig:CrystalImages}(b).

\subsection{Crystal Morphology \& Compositional Analysis}

\begin{figure}[htb]
    \centering
    \includegraphics[width=\columnwidth]{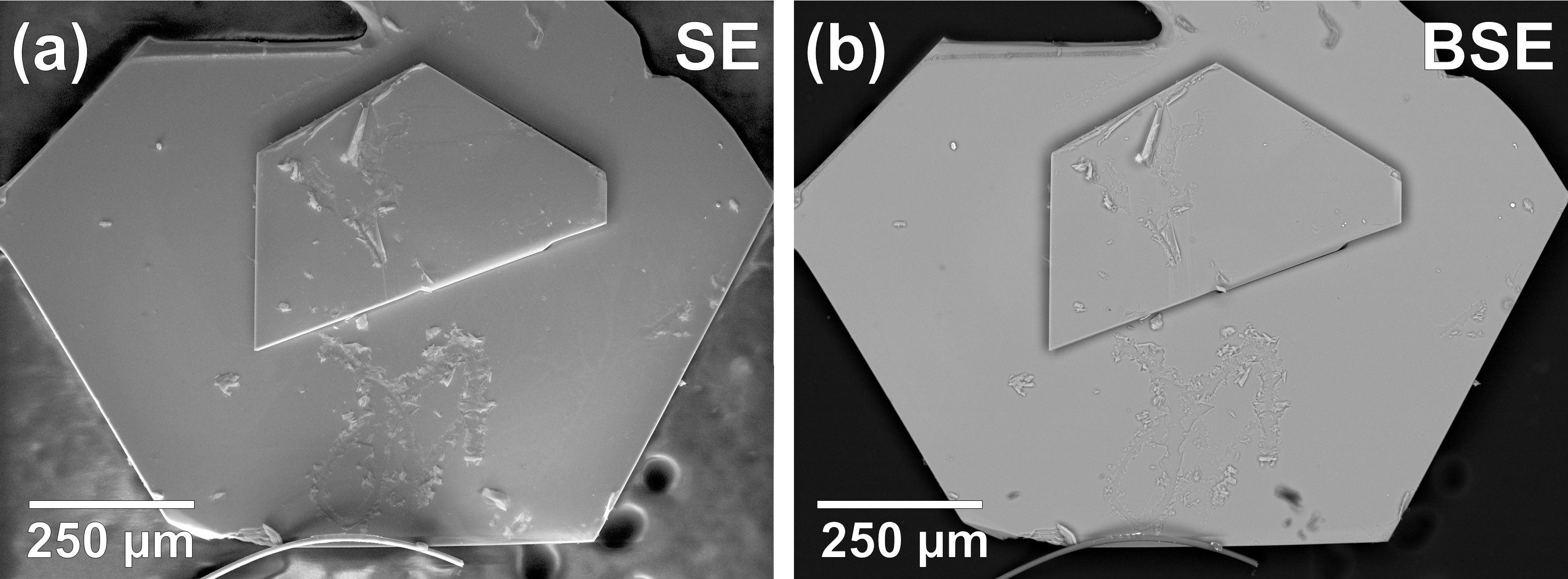}
    \caption{SEM images of a CuCrP$_2$S$_6$ crystal with (a) topographical contrast (using an SE detector) and (b) with chemical contrast (using a BSE detector).}
    \label{fig:SEM_SE_BSE}
\end{figure}

The crystal morphology was investigated in detail by SEM. Using a SE detector (\textit{i.e.} topographic contrast), flat surfaces and well-developed edge facets, as, \textit{e.g.}, illustrated in Fig.~\ref{fig:SEM_SE_BSE}(a), were resolved on our CuCrP$_2$S$_6$ crystals. The edge facets indicate a hexagonal crystal habitus. In the lower central part of Fig.~\ref{fig:SEM_SE_BSE}(a), the surface of the crystal was scratched and several layers peeled off, demonstrating the layered nature of the crystal with only weak van-der-Waals interactions between adjacent layers.

Using the chemical contrast mode (BSE detector) as shown in Fig.~\ref{fig:SEM_SE_BSE}(b), a homogeneous contrast of the crystal surface was detected. Comparing the SEM(SE) and SEM(BSE) images, the few spots of different contrast can be clearly attributed to particles on the crystal surface and not to intrinsic impurities in the crystal. This implies, subsequently, that the elemental composition is homogeneous over the whole crystal.

This is verified by measurements of the elemental composition by EDX on 30 spots on multiple crystals of the same crystal growth experiment. The mean elemental composition is Cu$_{1.05(3)}$Cr$_{1.04(2)}$P$_{2.03(1)}$S$_{5.88(4)}$, in good agreement with the expected composition of CuCrP$_2$S$_6$. Furthermore, the low standard deviations indicate a high homogeneity of the elemental composition over all studied crystals. Also, no significant amount of the transport agent iodine was found incorporated in any crystal.

\subsection{Exfoliation of as grown crystals}

\begin{figure*}[htb]
    \centering
    \includegraphics[width=0.7\linewidth]{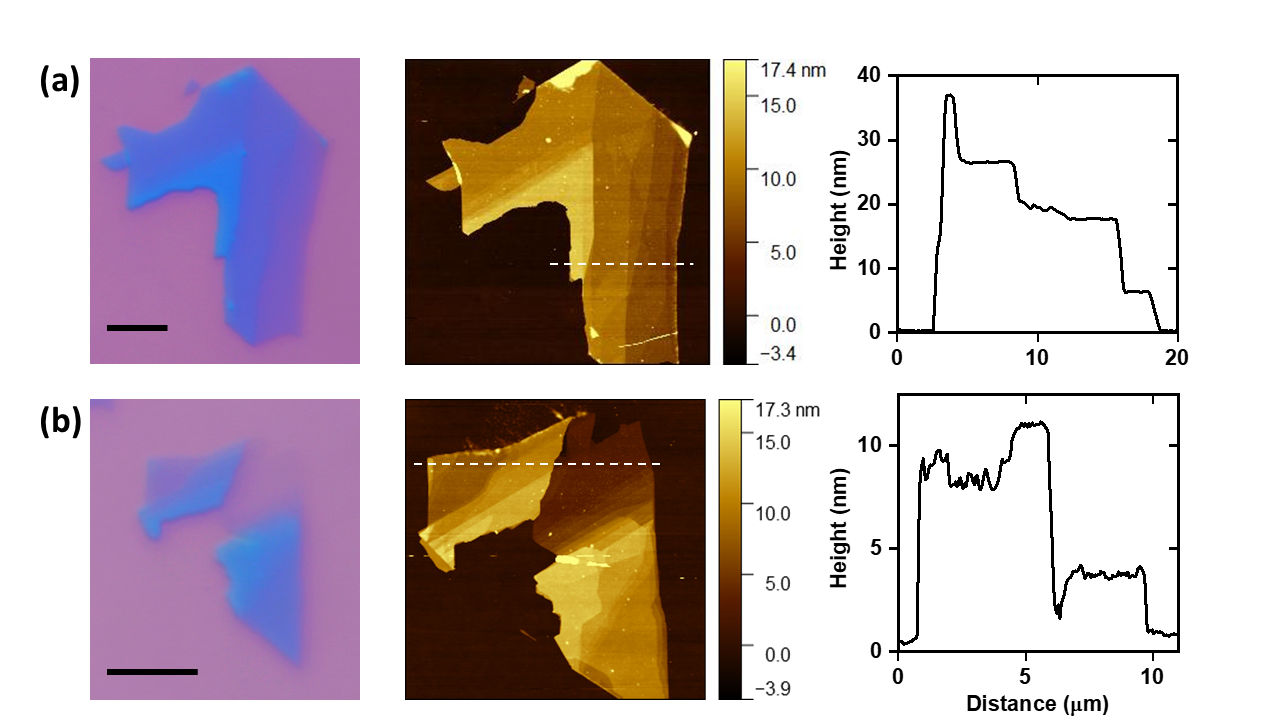}
    \caption{Exfoliated flakes of CuCrP$_2$S$_6$ on SiO$_2$ substrates. Left panel: optical image (scale-bar: 5$\mu$m). Center panel: atomic force microscope image of the exfoliated flake. Right panel: Height profile extracted from the AFM image  along the white dashed line in the center panel. (a): typical stepped flake with several flat areas, from 35~nm down to 6~nm; (b): exfoliated flake with a 4-layer step (2.8~nm).}
    \label{fig:EXFO}
\end{figure*}

As-grown CuCrP$_2$S$_6$ crystal were exfoliated using the standard scotch-tape technique on SiO$_2$ substrates. The exfoliation process yields many micrometer-scale crystallites, but also larger flat flakes in the tens of micrometer scale, with thicknesses ranging from about 50~nm down to a few layers (see Fig.~\ref{fig:EXFO}(a)). The thickness of the exfoliated flakes was characterized by atomic force microscopy (AFM). In particular, as shown in Fig.~\ref{fig:EXFO}(b), we obtained a 4-layer flake (2.8~nm thickness) with dimensions of about 3x3$\mu$m$^2$. The flakes do not show any optical sign of degradation in air over several weeks. The possibility to exfoliate large few-layers flakes of CuCrP$_2$S$_6$ enables future local measurements on the micro- and nanometer scale with optical techniques, such as Kerr microscopy, as well as its use in van-der-Waals heterostructures.

\subsection{Structural Analysis}

\begin{figure}[htb]
    \centering
    \includegraphics[width=\columnwidth]{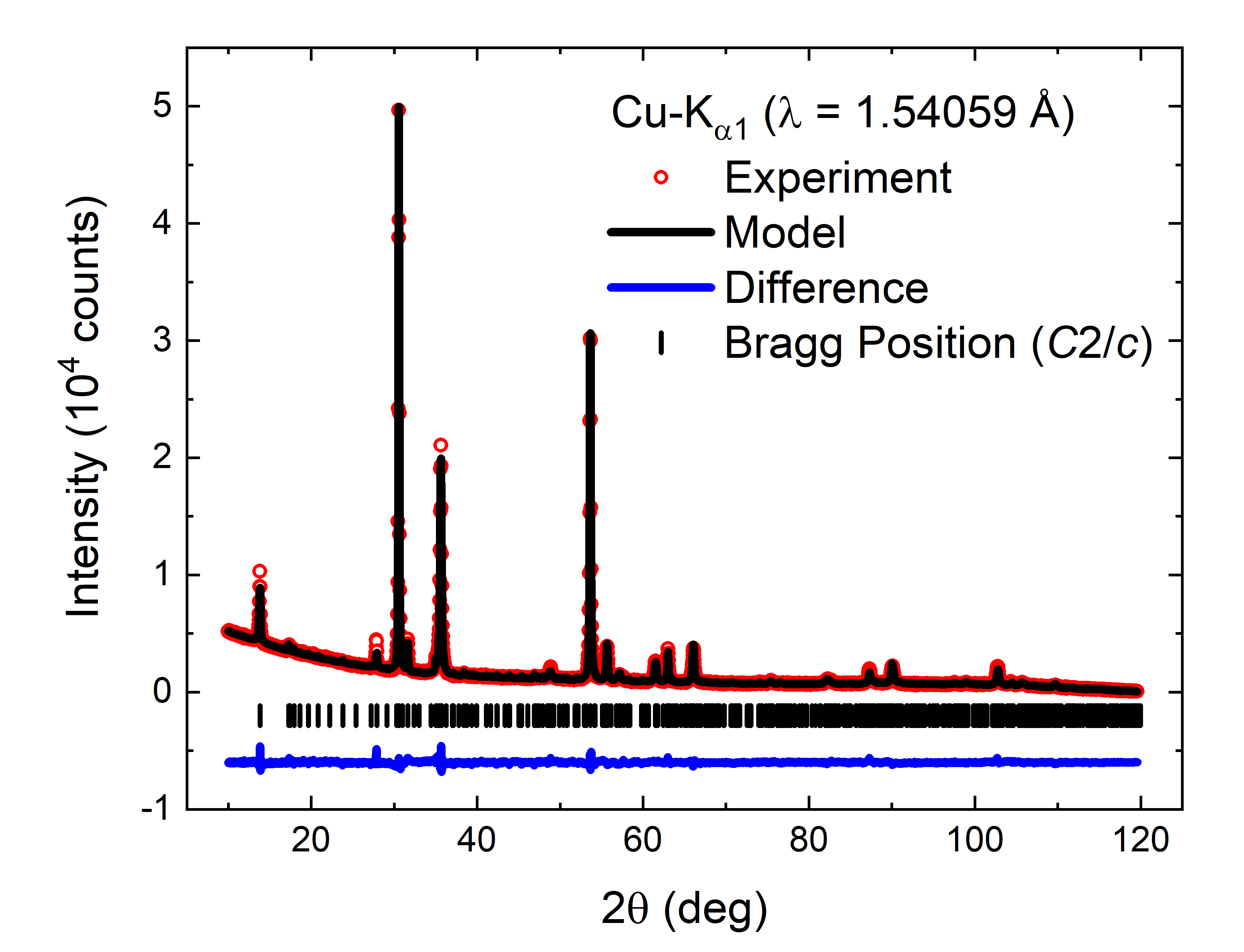}
    \caption{pXRD pattern of pulverized CuCrP$_2$S$_6$ crystals from Cu-K${_{\alpha_1}}$ radiation ($\lambda = 1.54059$\,\r{A}) compared to the calculated pattern based on the optimized structural model from Rietveld refinement.}
    \label{fig:pXRD_pattern}
\end{figure}

The measured pXRD pattern, shown in Fig.~\ref{fig:pXRD_pattern}, could be indexed in the monoclinic crystal system and was subsequently modeled by the Rietveld method starting from the crystal structure model in the monoclinic space group $C2/c$ proposed in Ref.~\cite{PColombet1982}.

The refined crystal structure model obtained by our Rietveld refinement exhibits zero occupation of the Cu1 site (Wyckoff site 8f) close to the plane shared by all Cr atoms in a layer, in agreement with other structural investigations of this compound~\cite{Susner2020}. Consequently, all Cu atoms are notably displaced away from the center of their surrounding S$_6$ octahedra (along $c^*$). The site containing all Cu atoms in our model is located close to the Cu2 site proposed in Ref.~\cite{PColombet1982} even after a refinement including all spatial coordinates of this position. Accordingly, the coordination environment of Cu in CuCrP$_2$S$_6$ is best described as trigonal planar rather than octahedral, which would naively be expected in similarity to the $M$ position in $M_2$P$_2$S$_6$. A summary and the reliability factors after the final iteration of the Rietveld analysis are shown in Table~\ref{tab:pXRD_Rietveld} and details about the refined structural model are given in Table~\ref{tab:refined_atomic_model}.

\begin{table}[htb]
 \caption{Summary and reliability factors of the Rietveld analysis of the pXRD pattern of CuCrP$_2$S$_6$.}
 \centering
 \begin{ruledtabular}
 \begin{tabular}{ld}
 \multicolumn{2}{l}{\textbf{Experiment \& Data Collection}} \\
 Temperature (K) & \multicolumn{1}{c}{293(2)} \\
 Radiation Type & \multicolumn{1}{c}{Cu-K$_{\alpha_1}$} \\
 Radiation Wavelength (\r{A}) & \multicolumn{1}{c}{1.54059} \\
 $\theta_\textrm{min}$ ($^\circ$) & 10.00 \\
 $\theta_\textrm{step}$ ($^\circ$) & 0.03 \\
 $\theta_\textrm{max}$ ($^\circ$) & 119.62 \\
 &\\
 \multicolumn{2}{l}{\textbf{Crystal Data}} \\
 Crystal System & \multicolumn{1}{c}{Monoclinic} \\
 Space Group & \multicolumn{1}{c}{$C2/c$} \\
 $a$ (\r{A}) & 5.9098(2) \\
 $b$ (\r{A}) & 10.2447(4) \\
 $c$ (\r{A}) & 13.3644(12) \\
 $\beta$ ($^\circ$) & 106.974(5) \\
 &\\
 \multicolumn{2}{l}{\textbf{Refinement}} \\
 Goodness-Of-Fit & 2.09 \\
 $R_\textrm{p}$ (\%) & 3.76 \\
 $wR_\textrm{p}$ (\%) & 5.25 \\
 $R_\textrm{F}$ (\%) & 10.72 \\
 \end{tabular}
 \end{ruledtabular}
 \label{tab:pXRD_Rietveld}
\end{table}

\begin{table*}[htb]
 \caption{Fractional atomic coordinates, occupancies and isotropic displacement parameters $U_\textrm{iso}$ of the structural model of CuCrP$_2$S$_6$ in $C2/c$ after Rietveld refinement. Estimated standard deviations are given in parentheses. Please note that the Cu1 site is empty and only listed here for comparison with the structure proposed by Colombet \textit{et al.}~\cite{PColombet1982}.}
 \centering
 \begin{ruledtabular}
 \begin{tabular}{cccccccc}
 \multirow{2}{*}{\textbf{Label}} & \multirow{2}{*}{\textbf{Type}} & \multirow{2}{*}{\textbf{Wyck}} & \multirow{2}{*}{$\boldsymbol{x}$} & \multirow{2}{*}{$\boldsymbol{y}$} & \multirow{2}{*}{$\boldsymbol{z}$} & \textbf{Occ} & $\boldsymbol{U}_\textrm{iso}$ \\
  & & & & & & \textbf{(\%)} & \textbf{($\times 10^{-3} \textrm{\r{A}}^2$)} \\
 \textcolor{gray}{Cu1} & \textcolor{gray}{Cu} & \textcolor{gray}{$8f$} & \textcolor{gray}{0.4966}     & \textcolor{gray}{0.5020}     & \textcolor{gray}{0.2670}        & \textcolor{red}{0}   &  \textcolor{gray}{-}    \\
 Cu2 & Cu & $8f$ & 0.0460(12) & 0.0018(13) & 0.3275(7)     & 50  &  9(2) \\
 Cr1 & Cr & $4e$ & 0          & 0.3315(9)  & 0.25          & 100 & 18(2) \\
 P1  & P  & $8f$ & 0.037(2)   & 0.3350(13) & 0.8214(9)     & 100 & 61(3) \\
 S1  & S  & $8f$ & 0.2499(15) & 0.1741(15) & 0.3695(7)     & 100 & 27(2) \\
 S2  & S  & $8f$ & 0.2676(15) & 0.1756(13) & 0.8757(7)     & 100 & 18(2) \\
 S3  & S  & $8f$ & 0.7374(11) & 0.9975(14) & 0.3751(7)     & 100 & 15(2) \\
 \end{tabular}
 \end{ruledtabular}
 \label{tab:refined_atomic_model}
\end{table*}

It should be mentioned, that also a crystal structure model in the space group $C2/m$ with virtually the same lattice parameters and a random distribution of Cu and Cr on the $M$ site of the $M_2$P$_2$S$_6$ structure (analogue to FeNiP$_2$S$_6$, MnNiP$_2$S$_6$ and MnFeP$_2$S$_6$) yields a calculated pXRD pattern that is in reasonable agreement with our experimental pattern. This is according to expectation, as both $C2/m$ and $C2/c$ are part of the same Laue class ($2/m$) and both space groups exhibit the same centering. Consequently, the reflection conditions for both models are the same and they may only be distinguished based on slight differences in reflection intensities due to different structure factors. As the reflection intensities and shapes are additionally affected by stacking faults, twinning and preferred orientation of the crystallites in the powder, it is not possible to unambiguously determine which crystal structure model depicts reality more accurately based on the pXRD analysis alone.

Taking further information into account, however, the crystal structure model in $C2/c$ is more likely. On one hand, the diffraction pattern from single crystal X-ray diffraction experiments, which allow for a more precise analysis of the structure factor than pXRD, also show a better agreement with the structure in $C2/c$ than with the one in $C2/m$. On the other hand, our magnetic analysis, that is discussed in detail hereafter, implies the oxidation states of Cu$^{1+}$Cr$^{3+}$[P$_2$S$_6$]$^{4-}$ rather than Cu$^{2+}$Cr$^{2+}$[P$_2$S$_6$]$^{4-}$. As reported in Ref.~\cite{RBrec1986}, the minimization of Coulomb repulsion between the differently charged $M^{1+}$ and $M'^{3+}$ ions in $MM'$P$_2$S$_6$ favors an alternating arrangement of these cations (as, \textit{e.g.}, also reported for AgInP$_2$S$_6$ \cite{RBrec1986}). For CuCrP$_2$S$_6$, such a charge-induced alternating arrangement is only ensured by the crystal structure model in $C2/c$. If, in contrast, Cu and Cr would have the same oxidation state of 2+, a random distribution of these ions on the $M$ sites on the honeycomb lattice would be rather expected, as, \textit{e.g.}, observed for FeNiP$_2$S$_6$ and other homocharge substituted $M_2$P$_2$S$_6$ compounds \cite{Selter2021,Shemerliuk2021,TMasubuchi2008}.

\subsection{Quantum Chemical Calculations}

Our material model was devised as a finite fragment consisting of a CrS$_6$ octahedron and the adjacent six P ions plus 12 additional S anions around those P sites. The crystalline environment was described as a large array of point charges that reproduces the crystalline Madelung field within the cluster region; for this purpose we used the {\sc ewald} 
package \cite{Klintenberg_et_al,Derenzo_et_al}. We initiated the quantum chemical investigation at the level of complete active space self-consistent field (CASSCF) calculations \cite{qc_book_00}; an active space defined by the five 3$d$ orbitals of the central Cr atom was employed. Next, we correlated the Cr 3$d$ electrons and 3$p$ electrons of the central-octahedron S atoms in multireference configuration-interaction (MRCI) calculations with single and double excitations \cite{qc_book_00}. Finally, spin-orbit couplings were accounted for according to the procedure discussed in Ref.~\cite{SOC_molpro}. We used all-electron triple-$\zeta$ quality basis sets for the Cr ion \cite{Balabanov_Cr} and S ligands \cite{Dunning_S} 
of the central CrS$_6$ octahedron along with large-core pseudopotentials for the six P and 12 S nearest neighbors \cite{Igel-Mann_et_al}.

\begin{table}[htb]
\caption{ 
CASSCF and MRCI relative energies (eV) for the Cr$^{3+}$ 3$d^3$ multiplet structure in CuCrP$_2$S$_6$. 
Notations as in $O_h$ symmetry are used, although the actual point group symmetry is lower.
}
\begin{ruledtabular}
\begin{tabular}{l c c }

\textbf{3$d^3$ splittings}       &\textbf{CASSCF}    &\textbf{MRCI}         \\

$^4\!A_{2g}$ ($t^3_{2g}$)        &0.00                 &0.00                 \\
$^4\!T_{2g}$ ($t^2_{2g}e^1_g$)   &1.65; 1.71; 1.78     &1.76; 1.83; 1.90     \\
$^2\!E_{g}$  ($t^3_{2g}$)        &2.30; 2.31           &2.19; 2.20           \\
$^2\!T_{1g}$ ($t^3_{2g}$)        &2.41; 2.41; 2.48     &2.29; 2.29; 2.35     \\
$^4\!T_{1g}$ ($t^2_{2g}e^1_g$)   &2.54; 2.73; 2.82     &2.61; 2.79; 2.91     \\
$^2\!T_{2g}$ ($t^3_{2g}$)        &3.19; 3.32; 3.33     &2.99; 3.14; 3.14     \\
\end{tabular}
\end{ruledtabular}
\label{tab:excited_energy}
\end{table}

CASSCF and MRCI data for the Cr$^{3+}$ 3$d^3$ electronic structure are provided in Table\;\ref{tab:excited_energy}. This set of results corresponds to having two of the Cu$^+$ nearest neighbors on one side of the Cr-ion honeycomb plane, the third one on the other side; changing the positions of these Cu$^+$ cations such that they are all on the same side of the magnetic layer has only minor effects on the computed $d$-$d$ excitation energies, with variations of not more than $\sim$0.05 eV for the latter. For both geometries, the splittings induced by deviations from $O_h$ symmetry are moderate (see Table\;\ref{tab:excited_energy}).

The MRCI treatment brings sizable corrections to the CASSCF relative energies, of up to 0.2 eV. The $^4\!A_{2g}$--$^4\!T_{2g}$ gap, $\sim$1.8 eV, indicates the size of the crystal-field $t_{2g}$--$e_g$ splitting. By including spin-orbit interactions in the calculations \cite{SOC_molpro}, we find a rather small ground-state zero-field splitting of 0.1 meV (not shown in Table\;\ref{tab:excited_energy}).

\subsection{Magnetization}

\begin{figure*}[htb]
    \centering
    \includegraphics[width=\textwidth]{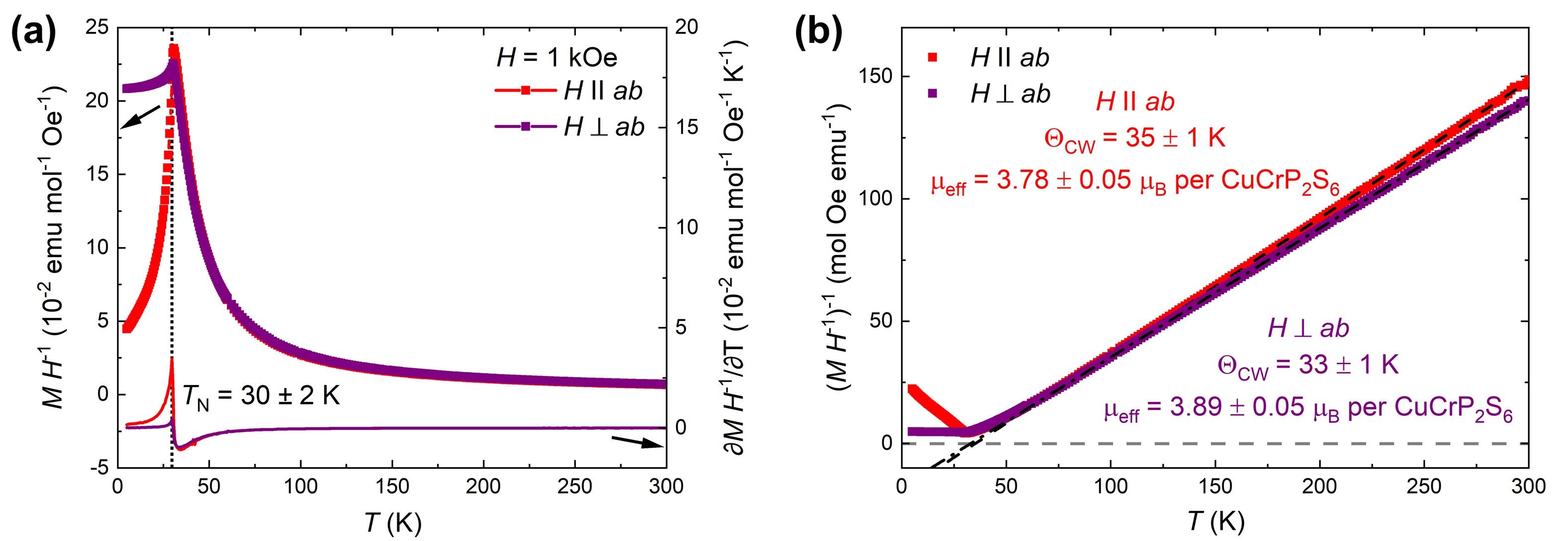}
    \caption{(a) Thermal evolution of the normalized magnetization $MH^{-1}(T)$ and its first derivative for $H = 1$\,kOe applied parallel and perpendicular to the crystallographic $ab$ plane. $T_\textrm{N}$ is indicated by the black dotted line. (b) Inverse of the normalized magnetization as function of temperature $(MH^{-1})^{-1}(T)$. The black dashed lines are linear regressions of the temperature regime 100--300\,K.}
    \label{fig:MH-1T_CWFit}
\end{figure*}

The normalized magnetization of CuCrP$_2$S$_6$ as function of temperature $MH^{-1}(T)$ is shown in Fig.~\ref{fig:MH-1T_CWFit}(a) for a magnetic field of 1\,kOe applied parallel and perpendicular to the $ab$ planes (\textit{i.e.} parallel to a random direction in the $ab$ plane and parallel to $c^*$, respectively). Additionally, on the bottom of Fig.~\ref{fig:MH-1T_CWFit}(a) the first derivative of $MH^{-1}(T)$ is shown.

From 300\,K towards lower temperatures, the normalized magnetization monotonically increases. A maximum in $MH^{-1}(T)$ is observed at $\sim 31$\,K together with an inflection point at slightly lower temperature, signalling a transition into an antiferromagnetically ordered state with the transition temperature $T_\textrm{N} = 30 \pm 2$\,K, in agreement with literature \cite{PColombet1982,WKleemann2011}. The normalized magnetization becomes strongly anisotropic below the maximum.

The inverse of the normalized magnetization of CuCrP$_2$S$_6$ as function of temperature is shown in Fig.~\ref{fig:MH-1T_CWFit}(b). At high temperatures a linear evolution is observed in agreement with the Curie-Weiss law. Our Curie-Weiss analysis in the temperature regime 100--300\,K yields $\Theta_\textrm{CW} = 35 \pm 1$\,K and $\mu_\textrm{eff} = 3.78 \pm 0.05\,\mu_\textrm{B}$~per~formula unit (f.u.) for $H \parallel ab$ and $\Theta_\textrm{CW} = 33 \pm 1$\,K and $\mu_\textrm{eff} = 3.89 \pm 0.05\,\mu_\textrm{B}$~per~f.u. for $H \perp ab$, and thus no sizeable magnetic anisotropy in the paramagnetic state. Our fitted Curie-Weiss values are in good agreement with two earlier works (31.5 K \cite{PColombet1982} and 30.6 K \cite{WKleemann2011}), emphasizing the dominance of the ferromagnetic coupling competing with the smaller antiferromagnetic interlayer coupling in this compound.

Regarding $\mu_\textrm{eff}$, both values (for $H \parallel ab$ and $H \perp ab$) are close to the magnetic moment expected for Cr$^{3+}$, $\mu_\textrm{eff}($Cr$^{3+}) = 3.87\,\mu_\textrm{B}$, assuming the free electron $g$-factor. This verifies that the oxidation states are indeed Cu$^{1+}$Cr$^{3+}$[P$_2$S$_6$]$^{4-}$ rather than  Cu$^{2+}$Cr$^{2+}$[P$_2$S$_6$]$^{4-}$, for which a significantly larger magnetic moment per CuCrP$_2$S$_6$ would be expected in the paramagnetic state (theoretical values: $\mu_\textrm{eff}($Cr$^{2+}) = 4.90\,\mu_\textrm{B}$ and $\mu_\textrm{eff}($Cu$^{2+}) = 1.73\,\mu_\textrm{B}$).

\begin{figure*}[htb]
    \centering
    \includegraphics[width=\textwidth]{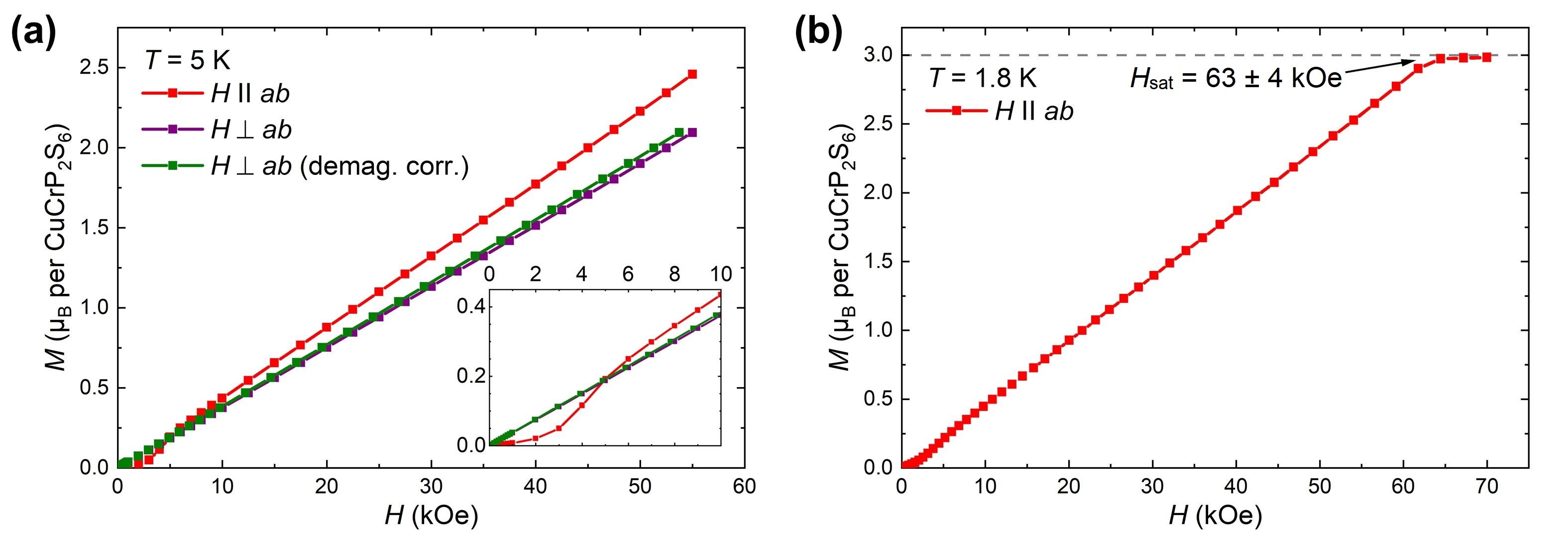}
    \caption{(a) Field dependent magnetization $M(H)$ at 5\,K with $H \leq 55$\,kOe applied parallel (red) and perpendicular (violet) to the crystallographic $ab$ plane. The green curve shows $M(H)$ for $H \perp ab$ after demagnetizing field correction as explained in the text. The inset shows a zoomed-in view of the low-field regime up to 10\,kOe. (b) $M(H)$ at 1.8\,K with $H \leq 70$\,kOe applied parallel to the $ab$ plane. The spin-only saturation moment of Cr$^{3+}$ (assuming $S = 3/2$ and $g = 2$) is marked by the grey dotted line.}
    \label{fig:MH}
\end{figure*}

The field dependence of the magnetization $M(H)$ of CuCrP$_2$S$_6$ at 5\,K and up to 55\,kOe is shown in Fig.~\ref{fig:MH}(a) for magnetic fields applied parallel and perpendicular to the $ab$ plane. The in-plane direction of the magnetic field is the same as for the thermal evolution discussed before. The influence of a demagnetizing field due to an anisotropic shape of the sample (\textit{i.e.} thin plate-like crystal) was estimated using the demagnetizing factors for an infinite thin platelet ($N_x = N_y = 0$ and $N_z = 1$)~\cite{JOsborn1945}. For $H \perp ab$, the field dependence of the magnetization is linear in agreement with Kleemann \textit{et al.}~\cite{WKleemann2011} and the demagnetizing field correction results in a slight increase of the slope. For $H \parallel ab$, the magnetization is not affected by this correction and a curved $M(H)$ behavior is observed at low fields centered around approximately 4\,kOe. Above approximately 10\,kOe, the field dependence for $H \parallel ab$ becomes linear at 5 K, however with a higher slope than for $H \perp ab$. The curvature at low fields resembles the behavior expected for a field-driven spin reorientation (spin-flop transition) which is expected for an antiferromagnetically ordered system for magnetic fields applied along the the spin direction. While this scenario generally follows the expectation for CuCrP$_2$S$_6$, Lai \textit{et al.}\cite{Lai2019} reported several features regarding this regime in the isothermal magnetization of CuCrP$_2$S$_6$ that are not in line with a usual spin-flop transition: (i) its appearance depends on the cooling process, (ii) it does not change as a function of field orientation in the $ab$ plane, while CuCrP$_2$S$_6$ is unidirectional in the $ab$ plane according to their calculations. Thus, they interpret the curvature at low fields for $H \parallel ab$ in the context of magnetoelectric coupling rather than a spin-flop transition, where the magnetoelastic coupling could result in a local effective field, leading to a pinning effect during the magnetization process that could be enhanced by a field-cooling procedure. If the origin of the curvature in the low-field regime of the isothermal magnetization is related to such a complex phenomenon will need to be established by future experiments.

Additionally, $M(H)$ measured at 1.8\,K and up to 70\,kOe parallel to the $ab$ plane is shown in Fig.~\ref{fig:MH}(b) indicating magnetic saturation above $H_\textrm{sat} = 63 \pm 4$\,kOe with a saturation magnetization of $M_\textrm{sat} \approx 3\,\mu_\textrm{B}$~per~CuCrP$_2$S$_6$. This saturation magnetization is in agreement with $S = 3/2$ and the free electron $g$-factor. As Cr$^{3+}$ in an octahedral crystal field has $S = 3/2$, this further supports the before mentioned oxidation states of Cu$^{1+}$Cr$^{3+}$[P$_2$S$_6$]$^{4-}$. Interestingly, the saturation field is relatively low for a layered system with an antiferromagnetic ground state. For example, the structurally related antiferromagnetic phosphorus sulfides, such as Ni$_2$P$_2$S$_6$ and Fe$_2$P$_2$S$_6$, exhibit a magnetization at 1.8\,K and 70\,kOe which is at least an order of magnitude smaller than the expected saturation value \cite{Selter2021}. This demonstrates that the spins in CuCrP$_2$S$_6$ are relatively easy to be polarized by the application of a magnetic field, supporting the scenario of a weak antiferromagnetic interlayer coupling together with a strong ferromagnetic intralayer coupling in this compound, in agreement with the positive values of $\Theta_\textrm{CW}$ mentioned before.

\subsection{Heat Capacity}

\begin{figure}[htb]
    \centering
    \includegraphics[width=\columnwidth]{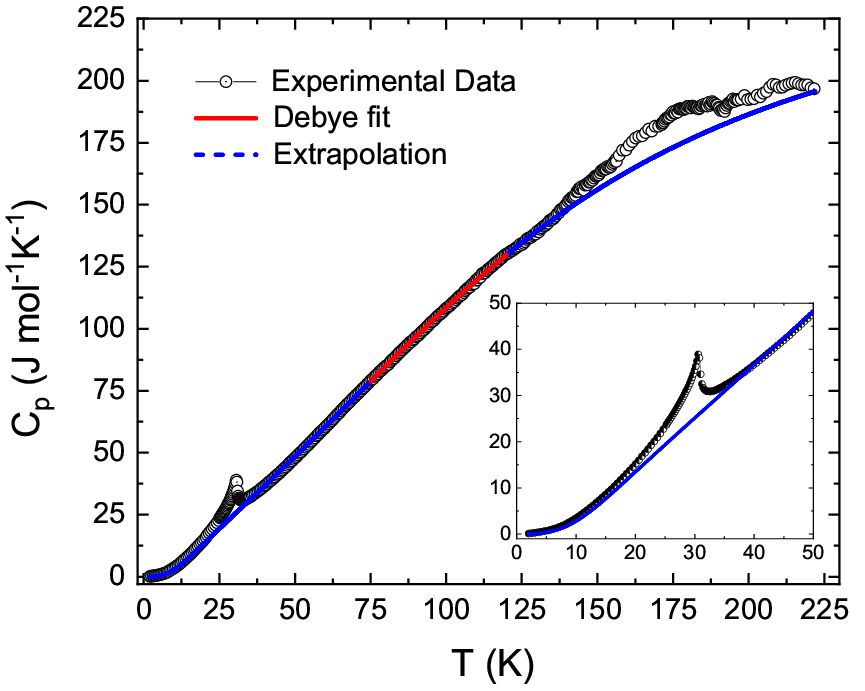}
    \caption{Temperature-dependent zero-field specific heat capacity of CuCrP$_2$S$_6$. The red line corresponds to the Debye model for the phononic contribution, while the blue line corresponds to the extrapolation of the obtained phonon function (Eq.~\ref{Debye}) for the whole temperature range. The inset shows a zoom into the low-temperature region with antiferromagnetic long-range ordering at $T_\textrm{N}$ = 31~K.}
    \label{fig:HC1}
\end{figure}

Fig. \ref{fig:HC1} shows the zero-field heat capacity $C_p$ of CuCrP$_2$S$_6$ as a function of temperature (black circles). As highlighted in the inset, a lambda-type second-order phase transition is observed at $T_\textrm{N} \sim 31$\,K due to antiferromagnetic ordering of the Cr$^{3+}$ ions, in good agreement with the magnetization curves.

Above $\sim 130$\,K, additional broad anomalies are present up to about 200-220~K. Such features are consistent with two transitions observed previously \cite{Susner2020,Moriya2005}, which are associated with structural changes accompanying antiferroelectric ordering triggered by the appearance of an ordered Cu sublattice. According to Ref.~\cite{Cajipea1996}, the first gradual transition around $T = 190$\,K from paralectric to an intermediate quasi-antipolar state has been interpreted in terms of random flipping dipoles in an otherwise antipolar phase, or static clusters of up or down disordered dipoles preceding a cooperative long-range order. A second transition around $T = 145$\,K marks the beginning of a fully antipolar state, which derives in one of the most interesting characteristics of CuCrP$_2$S$_6$ at low temperatures: the coexistence of antiferroelectric and antiferromagnetic lattices on different cation sites (Cu and Cr, respectively).

In order to obtain the magnetic contribution to the specific heat, the lattice contribution ($C_\textrm{ph}$) of CuCrP$_2$S$_6$ has to be removed from the total specific heat. The $C_\textrm{ph}$ contribution was estimated by fitting an intermediate range between the magnetic and structural transitions, corresponding to 75--120\,K (red line in Fig. \ref{fig:HC1}). The fit of the zero-field heat capacity was performed with the superposition of four Debye functions given by

\begin{equation}\label{Debye}
  C_\textrm{ph}= 9R\sum_{i=1}^{4}n_i\left(\frac{T}{\Theta_{\textrm{D}i}}\right)^3\int_{0}^{\Theta_{\textrm{D}i}/T}\frac{x^4e^x}{(e^x-1)^2}dx ,
\end{equation}

where $R$ is the universal gas constant, the index $i$ sums over the four different atom classes in the unit cell. n$_i$, and $\Theta_{\textrm{D}i}$~$(i=1,2,3,4)$ are the number of atoms and Debye temperatures for each $i^{th}$ atom class in the unit cell, respectively ($\sum_{i}n_i$ = 10). This multiple Debye fit allows a precise modeling of the specific heat in the low-temperature region (below the magnetic transition), and which is a different approach from an earlier report where a sum of one Debye function and four Einstein functions was used \cite{Moriya2005}. The obtained Debye temperatures are shown in Table \ref{tab:Cptable}.

\begin{table}[htb]
 \caption{Parameters obtained from $C_p$ modeling of CuCrP$_2$S$_6$.  $S_\textrm{mag}$ and $S_\textrm{str}$ denote the entropy release at the magnetic and structural phase transition in the shown temperature regions, respectively.}
 \centering
 \begin{ruledtabular}
 \begin{tabular}{lr}
 Debye Temperatures ($\Theta_\textrm{D}$) & 630 K, 365 K, 198 K, 92 K\\
 S$_\textrm{mag}$ (2--100\,K) & $3.67\pm0.22$ J\,mol$^{-1}$\,K$^{-1}$ \\
 S$_\textrm{str}$ (120--220\,K) & $3.82\pm0.1$ J\,mol$^{-1}$\,K$^{-1}$ \\
 \end{tabular}
 \end{ruledtabular}
 \label{tab:Cptable}
\end{table}

\begin{figure}[htb]
    \centering
    \includegraphics[width=\columnwidth]{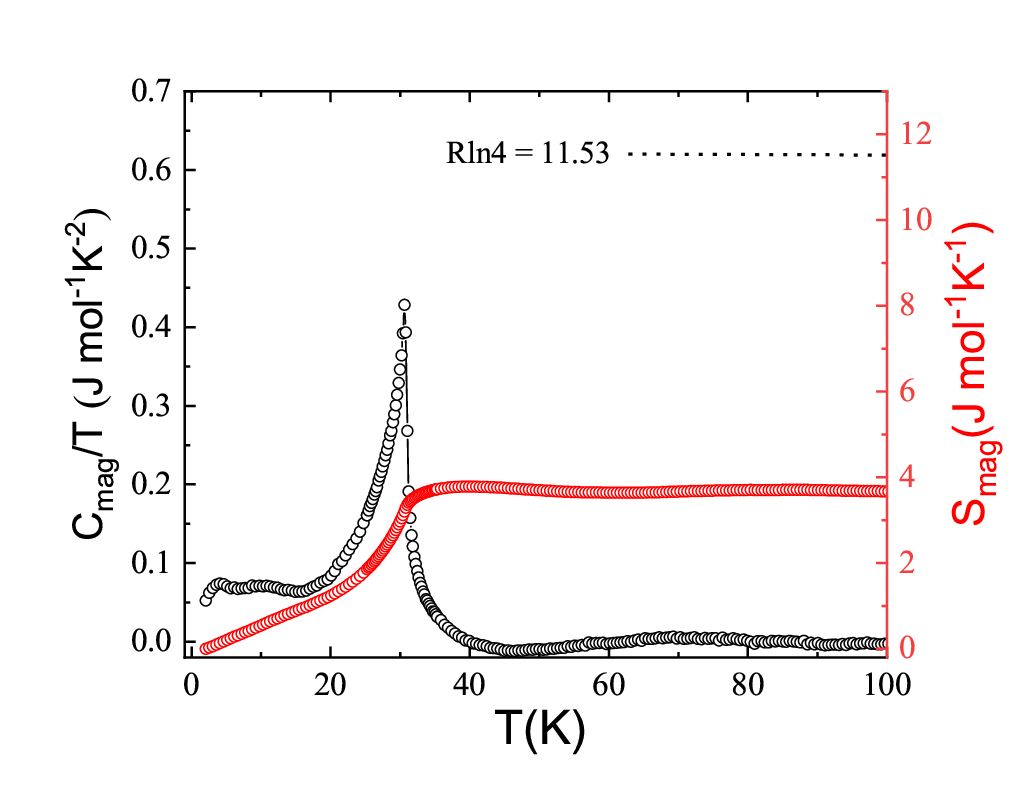}
    \caption{Zero-field magnetic specific heat of CuCrP$_2$S$_6$ plotted as $C_\textrm{mag}/T$ vs $T$ (left scale) and the corresponding magnetic entropy $S_\textrm{mag}(T)$ (right scale).}
    \label{fig:HC2}
\end{figure}

The estimation of the phonon contribution was obtained by extrapolating the obtained function to the whole temperature range (2--220\,K), shown in blue in Fig. \ref{fig:HC1} together with the corresponding magnetic entropy obtained by integration of $C_{mag}/T$ shown in Fig. \ref{fig:HC2}. An entropy release corresponding to the magnetic transition is observed up to $T$ $\sim$ 40~K,  followed by a plateau with $S_{mag}$ = $3.67 \pm 0.22$\,J\,mol$^{-1}$\,K$^{-1}$. This is in agreement with observations by Moriya \textit{et al.}~\cite{Moriya2005} with a slightly higher value of $S_{mag}$ = 4.10\,J\,mol$^{-1}$\,K$^{-1}$. Since for Cr$^{3+}$ with $S$ = 3/2 a magnetic entropy of $S_\textrm{mag} = R \ln(2S+1) = 11.5$\,J\,mol$^{-1}$\,K$^{-1}$ is expected, the missing entropy could possibly be ascribed to either (i) a non-perfect estimation of the lattice contribution due to the presence of short-range magnetic correlations well above the magnetic transition (typical for low-dimensional compounds and/or possibly related to the structural transitions at high temperature resulting from a coupling of spin and lattice degree of freedoms), and which are not taken into account in our modeling performed in the temperature regime 75-120~K, or (ii) a non-perfect structural 3D modeling for this quasi-2D van-der-Waals magnet. Note, that although a 2D Debye modeling substantially increased $S_{mag}$ towards the expected value, the low-temperature region could not be fitted in a satisfying manner, resulting in systematic unphysical deviations from our data. This possibly hints at neither purely 2D nor 3D phononic modes in CuCrP$_2$S$_6$ and calls for theoretical calculations based on the actual structural refinement. 

\begin{figure}[htb]
    \centering
    \includegraphics[width=\columnwidth]{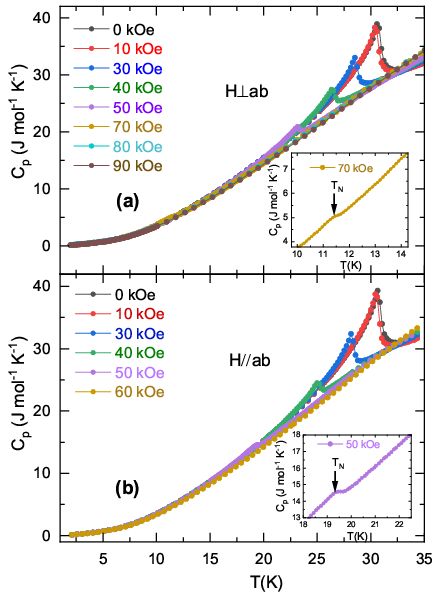}
    \caption{Temperature and field dependence of the specific heat for CuCrP$_2$S$_6$ (a) for fields applied  perpendicular to the $ab$ plane. The inset shows specific heat for 70 kOe. (b) For fields applied parallel to the $ab$ plane. The inset shows specific heat for 50 kOe}
    \label{fig:HC3}
\end{figure}

In order to probe the suppression of the long-range antiferromagnetic order, specific heat capacity studies were conducted in fields up to 9~T. Fig. \ref{fig:HC3} shows the temperature dependence of the specific heat in magnetic fields applied parallel and perpendicular to the $ab$ plane. For $H \perp ab$ the transition temperature shifts to lower temperature, while the magnitude of the peak decreases with increasing magnetic field, in line with the antiferromagnetic nature of the phase transition at $T_\textrm{N}$. The transition is completely suppressed for a magnetic field $\mu_0 H = 8$\,T. A similar behavior is detected for $H \parallel ab$, however, with a slightly lower critical field of about 60 kOe. Our results are in line with theoretical calculations \cite{Lai2019} which have found the $a$ axis to be the direction of easy magnetization.

\section{Summary \& Outlook}

In summary, optimized single crystals of the quaternary compound CuCrP$_2$S$_6$ were grown by CVT using iodine as transport agent. The pXRD pattern from pulverized crystals exhibits reflections that are overall in agreement with the reported structure in the space group $C2/c$~\cite{PColombet1982,PColombet1983}. A displacement of all Cu atoms along the $c^*$ direction away from the center towards the top and bottom faces of the surrounding S$_6$ octahedra was observed in our analysis.

Based on this structural model, quantum chemical calculations yield details of the Cr$^{3+}$ $3d^3$ multiplet structure with a octahedral crystal field splitting of $\sim 1.8$\,eV. Since the excitation energies obtained from our calculations define a rather closely spaced set of levels, we expect that a rather broad spectral feature will be found at 1.5--3.5 eV energy loss in future Cr $L$-edge resonant inelastic X-ray scattering (RIXS) experiments on CuCrP$_2$S$_6$.

In addition to the structural order of the Cu sublattice, magnetic ordering of the Cr sublattice has been observed in our single crystals. The thermal evolution of the normalized magnetization $MH^{-1}(T)$ of our CuCrP$_2$S$_6$ crystals shows an antiferromagnetic ground state, while our Curie Weiss analysis in a wide range of temperature in the paramagnetic state yields positive values of $\Theta_\textrm{CW}$, confirming dominant ferromagnetic interactions. 

At an external field of approximately 60\,kOe, magnetic saturation sets in for $H \parallel ab$. The low saturation field compared to other layered antiferromagnets implies that the spin system can easily be polarized, in agreement with dominant ferromagnetic (intralayer) interactions as implied by $\Theta_\textrm{CW}$. Both the magnetic saturation moment $M_\textrm{sat} = 3\,\mu_\textrm{B}$~per~f.u. and the effective moment $\mu_\textrm{eff} \approx 3.8\,\mu_\textrm{B}$~per~f.u. obtained from the Curie-Weiss analysis demonstrate that the magnetic ion in the system has $S = 3/2$ as expected for Cr$^{3+}$, which agrees with the oxidation states of Cu$^{1+}$Cr$^{3+}$P$_2$S$_6$.

Zero field specific heat measurements confirm the magnetic ordering temperature and the entropy release due to the magnetic phase transition was obtained as $S_\textrm{mag} = 3.67 \pm 0.22$\,J\,mol$^{-1}$\,K$^{-1}$. At higher temperatures between 140-200~K additional broad anomalies associated with structural changes accompanying antiferroelectric ordering are detected. Specific heat measurements in magnetic fields applied $\parallel ab$ and $\perp ab$ show that the magnetic order is suppressed in an external magnetic field of about 6\,T (8\,T) applied parallel (perpendicular) to the $ab$ plane.

Several characteristic features of the magnetism in CuCrP$_2$S$_6$ resemble similarities to CrCl$_3$~\cite{MMcGuire2017,JZeisner2020}. Both compounds exhibit an antiferromagnetic groundstate with a pronounced increase of the magnetization approaching the antiferromagnetic ordering temperature from the high temperature side as well as Curie-Weiss temperatures, which imply dominantly ferromagnetic interactions. For the latter compound, the increase of magnetization is attributed to the onset of 2D ferromagnetic order before at lower temperatures adjacent layers order antiferromagnetically. Although many magnetic characteristics appear to be similar, for CuCrP$_2$S$_6$ no second anomaly could be observed in our specific heat measurements, which implies that the increase in magnetization is not linked to the existence of an (ferromagnetically) ordered state in between the antiferromagnetic and paramagnetic states. Despite a similar magnetic ground state/spin structure of both compounds with dominant ferromagnetic intralayer coupling and antiferromagnetic coupling between such ferromagnetic layers, as detected in early neutron-diffraction experiments of CuCrP$_2$S$_6$ at $T$ = 11 K ($<$ $T_\textrm{N}$)~\cite{Maisonneuve1995}, it seems as if the character/strength of the ferromagnetic 2D correlations are distinctly different for both compounds.

In the context of a magnetic structure with ferromagnetic layers which are weakly antiferromagnetically coupled to each other~\cite{Maisonneuve1995}, it will be interesting to investigate the magnetic properties of CuCrP$_2$S$_6$ while thinning down the sample towards few layers or even the monolayer limit. As the antiferromagnetic contribution is related to interactions between layers, it is plausible that it will be weakened by reducing the amount of layers in the sample. Potentially in the monolayer limit, the material may acts as a 2D ferromagnet down to lowest temperatures as no interlayer antiferromagnetic interactions are present anymore. Our results on the exfoliation of thin flakes of CuCrP$_2$S$_6$ are the initial step towards such magnetic investigations in the future.

\section*{Credit author statement}
S.S., Y.S. and S.A. grew the crystals and characterized them. Ö.B. and L.V. have exfoliated the as-grown crystals into four layers. P.B. and L.H. performed the quantum chemistry calculations. K.K.B., L.T.C. and A.U.B.W. did the specific heat measurements. S.A. and B.B. initiated the work. All authors contributed to the analysis of the results. S.S. and S.A. wrote the publication with the input from all the co-authors.
 
\section*{Declaration of competing interest}
The authors declare that they have no known competing financial interests or personal relationships that could have appeared to influence the work reported in this paper.

\section*{Acknowledgements} This work is supported by the Deutsche Forschungsgemeinschaft (DFG) via Grant No.~DFG~A.S~523\textbackslash4-1. L.T.C. is funded by the DFG (project-id 456950766). A.U.B.W. and B.B. acknowledge financial support from the DFG through the collaborative research center SFB 1143 (project-id 247310070), and the W\"urzburg-Dresden Cluster of Excellence on Complexity and Topology in Quantum Matter -- \textit{ct.qmat} (EXC 2147, project-id 390858490). L.V. was supported by the Leibniz Association through the Leibniz Competition. P.B. and L.H. thank U.~Nitzsche for technical assistance and the German Research Foundation (project-id 441216021) for financial support.

\bibliography{literature}

\begin{thebibliography}{46}%
\makeatletter
\providecommand \@ifxundefined [1]{%
 \@ifx{#1\undefined}
}%
\providecommand \@ifnum [1]{%
 \ifnum #1\expandafter \@firstoftwo
 \else \expandafter \@secondoftwo
 \fi
}%
\providecommand \@ifx [1]{%
 \ifx #1\expandafter \@firstoftwo
 \else \expandafter \@secondoftwo
 \fi
}%
\providecommand \natexlab [1]{#1}%
\providecommand \enquote  [1]{``#1''}%
\providecommand \bibnamefont  [1]{#1}%
\providecommand \bibfnamefont [1]{#1}%
\providecommand \citenamefont [1]{#1}%
\providecommand \href@noop [0]{\@secondoftwo}%
\providecommand \href [0]{\begingroup \@sanitize@url \@href}%
\providecommand \@href[1]{\@@startlink{#1}\@@href}%
\providecommand \@@href[1]{\endgroup#1\@@endlink}%
\providecommand \@sanitize@url [0]{\catcode `\\12\catcode `\$12\catcode
  `\&12\catcode `\#12\catcode `\^12\catcode `\_12\catcode `\%12\relax}%
\providecommand \@@startlink[1]{}%
\providecommand \@@endlink[0]{}%
\providecommand \url  [0]{\begingroup\@sanitize@url \@url }%
\providecommand \@url [1]{\endgroup\@href {#1}{\urlprefix }}%
\providecommand \urlprefix  [0]{URL }%
\providecommand \Eprint [0]{\href }%
\providecommand \doibase [0]{https://doi.org/}%
\providecommand \selectlanguage [0]{\@gobble}%
\providecommand \bibinfo  [0]{\@secondoftwo}%
\providecommand \bibfield  [0]{\@secondoftwo}%
\providecommand \translation [1]{[#1]}%
\providecommand \BibitemOpen [0]{}%
\providecommand \bibitemStop [0]{}%
\providecommand \bibitemNoStop [0]{.\EOS\space}%
\providecommand \EOS [0]{\spacefactor3000\relax}%
\providecommand \BibitemShut  [1]{\csname bibitem#1\endcsname}%
\let\auto@bib@innerbib\@empty
\bibitem [{\citenamefont {Geim}\ and\ \citenamefont
  {Grigorieva}(2013)}]{AGeim2013}%
  \BibitemOpen
  \bibfield  {author} {\bibinfo {author} {\bibfnamefont {A.~K.}\ \bibnamefont
  {Geim}}\ and\ \bibinfo {author} {\bibfnamefont {I.~V.}\ \bibnamefont
  {Grigorieva}},\ }\bibfield  {title} {\bibinfo {title} {Van der waals
  heterostructures},\ }\href {https://doi.org/10.1038/nature12385} {\bibfield
  {journal} {\bibinfo  {journal} {Nature}\ }\textbf {\bibinfo {volume} {499}},\
  \bibinfo {pages} {419} (\bibinfo {year} {2013})}\BibitemShut {NoStop}%
\bibitem [{\citenamefont {Gong}\ and\ \citenamefont {Zhang}(2019)}]{CGong2019}%
  \BibitemOpen
  \bibfield  {author} {\bibinfo {author} {\bibfnamefont {C.}~\bibnamefont
  {Gong}}\ and\ \bibinfo {author} {\bibfnamefont {X.}~\bibnamefont {Zhang}},\
  }\bibfield  {title} {\bibinfo {title} {Two-dimensional magnetic crystals and
  emergent heterostructure devices},\ }\href
  {https://doi.org/10.1126/science.aav4450} {\bibfield  {journal} {\bibinfo
  {journal} {Science}\ }\textbf {\bibinfo {volume} {363}},\ \bibinfo {pages}
  {eaav4450} (\bibinfo {year} {2019})}\BibitemShut {NoStop}%
\bibitem [{\citenamefont {Wang}\ \emph
  {et~al.}(2018{\natexlab{a}})\citenamefont {Wang}, \citenamefont {Sapkota},
  \citenamefont {Taniguchi}, \citenamefont {Watanabe}, \citenamefont
  {Mandrus},\ and\ \citenamefont {Morpurgo}}]{Wang2018}%
  \BibitemOpen
  \bibfield  {author} {\bibinfo {author} {\bibfnamefont {Z.}~\bibnamefont
  {Wang}}, \bibinfo {author} {\bibfnamefont {D.}~\bibnamefont {Sapkota}},
  \bibinfo {author} {\bibfnamefont {T.}~\bibnamefont {Taniguchi}}, \bibinfo
  {author} {\bibfnamefont {K.}~\bibnamefont {Watanabe}}, \bibinfo {author}
  {\bibfnamefont {D.}~\bibnamefont {Mandrus}},\ and\ \bibinfo {author}
  {\bibfnamefont {A.~F.}\ \bibnamefont {Morpurgo}},\ }\bibfield  {title}
  {\bibinfo {title} {Tunneling spin valves based on
  {Fe$_{3}$GeTe$_2$}/{hBN}/{Fe$_3$GeTe$_2$} van der waals heterostructures},\
  }\href {https://doi.org/10.1021/acs.nanolett.8b01278} {\bibfield  {journal}
  {\bibinfo  {journal} {Nano Letters}\ }\textbf {\bibinfo {volume} {18}},\
  \bibinfo {pages} {4303} (\bibinfo {year} {2018}{\natexlab{a}})}\BibitemShut
  {NoStop}%
\bibitem [{\citenamefont {Gibertini}\ \emph {et~al.}(2019)\citenamefont
  {Gibertini}, \citenamefont {Koperski}, \citenamefont {Morpurgo},\ and\
  \citenamefont {Novoselov}}]{Gibertini2019}%
  \BibitemOpen
  \bibfield  {author} {\bibinfo {author} {\bibfnamefont {M.}~\bibnamefont
  {Gibertini}}, \bibinfo {author} {\bibfnamefont {M.}~\bibnamefont {Koperski}},
  \bibinfo {author} {\bibfnamefont {A.~F.}\ \bibnamefont {Morpurgo}},\ and\
  \bibinfo {author} {\bibfnamefont {K.~S.}\ \bibnamefont {Novoselov}},\
  }\bibfield  {title} {\bibinfo {title} {Magnetic {2D} materials and
  heterostructures},\ }\href {https://doi.org/10.1038/s41565-019-0438-6}
  {\bibfield  {journal} {\bibinfo  {journal} {Nature Nanotechnology}\ }\textbf
  {\bibinfo {volume} {14}},\ \bibinfo {pages} {408} (\bibinfo {year}
  {2019})}\BibitemShut {NoStop}%
\bibitem [{\citenamefont {Zhong}\ \emph {et~al.}(2017)\citenamefont {Zhong},
  \citenamefont {Seyler}, \citenamefont {Linpeng}, \citenamefont {Cheng},
  \citenamefont {Sivadas}, \citenamefont {Huang}, \citenamefont {Schmidgall},
  \citenamefont {Taniguchi}, \citenamefont {Watanabe}, \citenamefont {McGuire},
  \citenamefont {Yao}, \citenamefont {Xiao}, \citenamefont {Fu},\ and\
  \citenamefont {Xu}}]{Zhong2017}%
  \BibitemOpen
  \bibfield  {author} {\bibinfo {author} {\bibfnamefont {D.}~\bibnamefont
  {Zhong}}, \bibinfo {author} {\bibfnamefont {K.~L.}\ \bibnamefont {Seyler}},
  \bibinfo {author} {\bibfnamefont {X.}~\bibnamefont {Linpeng}}, \bibinfo
  {author} {\bibfnamefont {R.}~\bibnamefont {Cheng}}, \bibinfo {author}
  {\bibfnamefont {N.}~\bibnamefont {Sivadas}}, \bibinfo {author} {\bibfnamefont
  {B.}~\bibnamefont {Huang}}, \bibinfo {author} {\bibfnamefont
  {E.}~\bibnamefont {Schmidgall}}, \bibinfo {author} {\bibfnamefont
  {T.}~\bibnamefont {Taniguchi}}, \bibinfo {author} {\bibfnamefont
  {K.}~\bibnamefont {Watanabe}}, \bibinfo {author} {\bibfnamefont {M.~A.}\
  \bibnamefont {McGuire}}, \bibinfo {author} {\bibfnamefont {W.}~\bibnamefont
  {Yao}}, \bibinfo {author} {\bibfnamefont {D.}~\bibnamefont {Xiao}}, \bibinfo
  {author} {\bibfnamefont {K.-M.~C.}\ \bibnamefont {Fu}},\ and\ \bibinfo
  {author} {\bibfnamefont {X.}~\bibnamefont {Xu}},\ }\bibfield  {title}
  {\bibinfo {title} {Van der waals engineering of ferromagnetic semiconductor
  heterostructures for spin and valleytronics},\ }\href
  {https://doi.org/10.1126/sciadv.1603113} {\bibfield  {journal} {\bibinfo
  {journal} {Science Advances}\ }\textbf {\bibinfo {volume} {3}},\ \bibinfo
  {pages} {e1603113} (\bibinfo {year} {2017})}\BibitemShut {NoStop}%
\bibitem [{\citenamefont {Casto}\ \emph {et~al.}(2015)\citenamefont {Casto},
  \citenamefont {Clune}, \citenamefont {Yokosuk}, \citenamefont {Musfeldt},
  \citenamefont {Williams}, \citenamefont {Zhuang}, \citenamefont {Lin},
  \citenamefont {Xiao}, \citenamefont {Hennig}, \citenamefont {Sales},
  \citenamefont {Yan},\ and\ \citenamefont {Mandrus}}]{LCasto2015}%
  \BibitemOpen
  \bibfield  {author} {\bibinfo {author} {\bibfnamefont {L.~D.}\ \bibnamefont
  {Casto}}, \bibinfo {author} {\bibfnamefont {A.~J.}\ \bibnamefont {Clune}},
  \bibinfo {author} {\bibfnamefont {M.~O.}\ \bibnamefont {Yokosuk}}, \bibinfo
  {author} {\bibfnamefont {J.~L.}\ \bibnamefont {Musfeldt}}, \bibinfo {author}
  {\bibfnamefont {T.~J.}\ \bibnamefont {Williams}}, \bibinfo {author}
  {\bibfnamefont {H.~L.}\ \bibnamefont {Zhuang}}, \bibinfo {author}
  {\bibfnamefont {M.-W.}\ \bibnamefont {Lin}}, \bibinfo {author} {\bibfnamefont
  {K.}~\bibnamefont {Xiao}}, \bibinfo {author} {\bibfnamefont {R.~G.}\
  \bibnamefont {Hennig}}, \bibinfo {author} {\bibfnamefont {B.~C.}\
  \bibnamefont {Sales}}, \bibinfo {author} {\bibfnamefont {J.-Q.}\ \bibnamefont
  {Yan}},\ and\ \bibinfo {author} {\bibfnamefont {D.}~\bibnamefont {Mandrus}},\
  }\bibfield  {title} {\bibinfo {title} {Strong spin-lattice coupling in
  {CrSiTe$_3$}},\ }\href {https://doi.org/10.1063/1.4914134} {\bibfield
  {journal} {\bibinfo  {journal} {{APL} Materials}\ }\textbf {\bibinfo {volume}
  {3}},\ \bibinfo {pages} {041515} (\bibinfo {year} {2015})}\BibitemShut
  {NoStop}%
\bibitem [{\citenamefont {Samarth}(2017)}]{Samarth2017}%
  \BibitemOpen
  \bibfield  {author} {\bibinfo {author} {\bibfnamefont {N.}~\bibnamefont
  {Samarth}},\ }\bibfield  {title} {\bibinfo {title} {Magnetism in flatland},\
  }\href {https://doi.org/10.1038/546216a} {\bibfield  {journal} {\bibinfo
  {journal} {Nature}\ }\textbf {\bibinfo {volume} {546}},\ \bibinfo {pages}
  {216} (\bibinfo {year} {2017})}\BibitemShut {NoStop}%
\bibitem [{\citenamefont {Baugher}\ \emph {et~al.}(2014)\citenamefont
  {Baugher}, \citenamefont {Churchill}, \citenamefont {Yang},\ and\
  \citenamefont {Jarillo-Herrero}}]{Baugher2014}%
  \BibitemOpen
  \bibfield  {author} {\bibinfo {author} {\bibfnamefont {B.~W.~H.}\
  \bibnamefont {Baugher}}, \bibinfo {author} {\bibfnamefont {H.~O.~H.}\
  \bibnamefont {Churchill}}, \bibinfo {author} {\bibfnamefont {Y.}~\bibnamefont
  {Yang}},\ and\ \bibinfo {author} {\bibfnamefont {P.}~\bibnamefont
  {Jarillo-Herrero}},\ }\bibfield  {title} {\bibinfo {title} {Optoelectronic
  devices based on electrically tunable p{\textendash}n diodes in a monolayer
  dichalcogenide},\ }\href {https://doi.org/10.1038/nnano.2014.25} {\bibfield
  {journal} {\bibinfo  {journal} {Nature Nanotechnology}\ }\textbf {\bibinfo
  {volume} {9}},\ \bibinfo {pages} {262} (\bibinfo {year} {2014})}\BibitemShut
  {NoStop}%
\bibitem [{\citenamefont {Li}\ \emph {et~al.}(2019)\citenamefont {Li},
  \citenamefont {Ruan},\ and\ \citenamefont {Zeng}}]{Li2019}%
  \BibitemOpen
  \bibfield  {author} {\bibinfo {author} {\bibfnamefont {H.}~\bibnamefont
  {Li}}, \bibinfo {author} {\bibfnamefont {S.}~\bibnamefont {Ruan}},\ and\
  \bibinfo {author} {\bibfnamefont {Y.-J.}\ \bibnamefont {Zeng}},\ }\bibfield
  {title} {\bibinfo {title} {Intrinsic van der waals magnetic materials from
  bulk to the {2D} limit: New frontiers of spintronics},\ }\href
  {https://doi.org/10.1002/adma.201900065} {\bibfield  {journal} {\bibinfo
  {journal} {Advanced Materials}\ }\textbf {\bibinfo {volume} {31}},\ \bibinfo
  {pages} {1900065} (\bibinfo {year} {2019})}\BibitemShut {NoStop}%
\bibitem [{\citenamefont {Fan}\ and\ \citenamefont {Wu}(2019)}]{Fan2019}%
  \BibitemOpen
  \bibfield  {author} {\bibinfo {author} {\bibfnamefont {F.~R.}\ \bibnamefont
  {Fan}}\ and\ \bibinfo {author} {\bibfnamefont {W.}~\bibnamefont {Wu}},\
  }\bibfield  {title} {\bibinfo {title} {Emerging devices based on
  two-dimensional monolayer materials for energy harvesting},\ }\href
  {https://doi.org/10.34133/2019/7367828} {\bibfield  {journal} {\bibinfo
  {journal} {Research}\ }\textbf {\bibinfo {volume} {2019}},\ \bibinfo {pages}
  {7367828} (\bibinfo {year} {2019})}\BibitemShut {NoStop}%
\bibitem [{\citenamefont {Gong}\ \emph {et~al.}(2017)\citenamefont {Gong},
  \citenamefont {Li}, \citenamefont {Li}, \citenamefont {Ji}, \citenamefont
  {Stern}, \citenamefont {Xia}, \citenamefont {Cao}, \citenamefont {Bao},
  \citenamefont {Wang}, \citenamefont {Wang}, \citenamefont {Qiu},
  \citenamefont {Cava}, \citenamefont {Louie}, \citenamefont {Xia},\ and\
  \citenamefont {Zhang}}]{CGong2017}%
  \BibitemOpen
  \bibfield  {author} {\bibinfo {author} {\bibfnamefont {C.}~\bibnamefont
  {Gong}}, \bibinfo {author} {\bibfnamefont {L.}~\bibnamefont {Li}}, \bibinfo
  {author} {\bibfnamefont {Z.}~\bibnamefont {Li}}, \bibinfo {author}
  {\bibfnamefont {H.}~\bibnamefont {Ji}}, \bibinfo {author} {\bibfnamefont
  {A.}~\bibnamefont {Stern}}, \bibinfo {author} {\bibfnamefont
  {Y.}~\bibnamefont {Xia}}, \bibinfo {author} {\bibfnamefont {T.}~\bibnamefont
  {Cao}}, \bibinfo {author} {\bibfnamefont {W.}~\bibnamefont {Bao}}, \bibinfo
  {author} {\bibfnamefont {C.}~\bibnamefont {Wang}}, \bibinfo {author}
  {\bibfnamefont {Y.}~\bibnamefont {Wang}}, \bibinfo {author} {\bibfnamefont
  {Z.~Q.}\ \bibnamefont {Qiu}}, \bibinfo {author} {\bibfnamefont {R.~J.}\
  \bibnamefont {Cava}}, \bibinfo {author} {\bibfnamefont {S.~G.}\ \bibnamefont
  {Louie}}, \bibinfo {author} {\bibfnamefont {J.}~\bibnamefont {Xia}},\ and\
  \bibinfo {author} {\bibfnamefont {X.}~\bibnamefont {Zhang}},\ }\bibfield
  {title} {\bibinfo {title} {Discovery of intrinsic ferromagnetism in
  two-dimensional van der waals crystals},\ }\href
  {https://doi.org/10.1038/nature22060} {\bibfield  {journal} {\bibinfo
  {journal} {Nature}\ }\textbf {\bibinfo {volume} {546}},\ \bibinfo {pages}
  {265} (\bibinfo {year} {2017})}\BibitemShut {NoStop}%
\bibitem [{\citenamefont {Huang}\ \emph {et~al.}(2017)\citenamefont {Huang},
  \citenamefont {Clark}, \citenamefont {Navarro-Moratalla}, \citenamefont
  {Klein}, \citenamefont {Cheng}, \citenamefont {Seyler}, \citenamefont
  {Zhong}, \citenamefont {Schmidgall}, \citenamefont {McGuire}, \citenamefont
  {Cobden}, \citenamefont {Yao}, \citenamefont {Xiao}, \citenamefont
  {Jarillo-Herrero},\ and\ \citenamefont {Xu}}]{BHuang2017}%
  \BibitemOpen
  \bibfield  {author} {\bibinfo {author} {\bibfnamefont {B.}~\bibnamefont
  {Huang}}, \bibinfo {author} {\bibfnamefont {G.}~\bibnamefont {Clark}},
  \bibinfo {author} {\bibfnamefont {E.}~\bibnamefont {Navarro-Moratalla}},
  \bibinfo {author} {\bibfnamefont {D.~R.}\ \bibnamefont {Klein}}, \bibinfo
  {author} {\bibfnamefont {R.}~\bibnamefont {Cheng}}, \bibinfo {author}
  {\bibfnamefont {K.~L.}\ \bibnamefont {Seyler}}, \bibinfo {author}
  {\bibfnamefont {D.}~\bibnamefont {Zhong}}, \bibinfo {author} {\bibfnamefont
  {E.}~\bibnamefont {Schmidgall}}, \bibinfo {author} {\bibfnamefont {M.~A.}\
  \bibnamefont {McGuire}}, \bibinfo {author} {\bibfnamefont {D.~H.}\
  \bibnamefont {Cobden}}, \bibinfo {author} {\bibfnamefont {W.}~\bibnamefont
  {Yao}}, \bibinfo {author} {\bibfnamefont {D.}~\bibnamefont {Xiao}}, \bibinfo
  {author} {\bibfnamefont {P.}~\bibnamefont {Jarillo-Herrero}},\ and\ \bibinfo
  {author} {\bibfnamefont {X.}~\bibnamefont {Xu}},\ }\bibfield  {title}
  {\bibinfo {title} {Layer-dependent ferromagnetism in a van der waals crystal
  down to the monolayer limit},\ }\href {https://doi.org/10.1038/nature22391}
  {\bibfield  {journal} {\bibinfo  {journal} {Nature}\ }\textbf {\bibinfo
  {volume} {546}},\ \bibinfo {pages} {270} (\bibinfo {year}
  {2017})}\BibitemShut {NoStop}%
\bibitem [{\citenamefont {Song}\ \emph {et~al.}(2019)\citenamefont {Song},
  \citenamefont {Tu}, \citenamefont {Carnahan}, \citenamefont {Cai},
  \citenamefont {Taniguchi}, \citenamefont {Watanabe}, \citenamefont {McGuire},
  \citenamefont {Cobden}, \citenamefont {Xiao}, \citenamefont {Yao},\ and\
  \citenamefont {Xu}}]{Song2019}%
  \BibitemOpen
  \bibfield  {author} {\bibinfo {author} {\bibfnamefont {T.}~\bibnamefont
  {Song}}, \bibinfo {author} {\bibfnamefont {M.~W.-Y.}\ \bibnamefont {Tu}},
  \bibinfo {author} {\bibfnamefont {C.}~\bibnamefont {Carnahan}}, \bibinfo
  {author} {\bibfnamefont {X.}~\bibnamefont {Cai}}, \bibinfo {author}
  {\bibfnamefont {T.}~\bibnamefont {Taniguchi}}, \bibinfo {author}
  {\bibfnamefont {K.}~\bibnamefont {Watanabe}}, \bibinfo {author}
  {\bibfnamefont {M.~A.}\ \bibnamefont {McGuire}}, \bibinfo {author}
  {\bibfnamefont {D.~H.}\ \bibnamefont {Cobden}}, \bibinfo {author}
  {\bibfnamefont {D.}~\bibnamefont {Xiao}}, \bibinfo {author} {\bibfnamefont
  {W.}~\bibnamefont {Yao}},\ and\ \bibinfo {author} {\bibfnamefont
  {X.}~\bibnamefont {Xu}},\ }\bibfield  {title} {\bibinfo {title} {Voltage
  control of a van der waals spin-filter magnetic tunnel junction},\ }\href
  {https://doi.org/10.1021/acs.nanolett.8b04160} {\bibfield  {journal}
  {\bibinfo  {journal} {Nano Letters}\ }\textbf {\bibinfo {volume} {19}},\
  \bibinfo {pages} {915} (\bibinfo {year} {2019})}\BibitemShut {NoStop}%
\bibitem [{\citenamefont {Wang}\ \emph
  {et~al.}(2018{\natexlab{b}})\citenamefont {Wang}, \citenamefont
  {Guti{\'{e}}rrez-Lezama}, \citenamefont {Ubrig}, \citenamefont {Kroner},
  \citenamefont {Gibertini}, \citenamefont {Taniguchi}, \citenamefont
  {Watanabe}, \citenamefont {Imamo{\u{g}}lu}, \citenamefont {Giannini},\ and\
  \citenamefont {Morpurgo}}]{Wang2018a}%
  \BibitemOpen
  \bibfield  {author} {\bibinfo {author} {\bibfnamefont {Z.}~\bibnamefont
  {Wang}}, \bibinfo {author} {\bibfnamefont {I.}~\bibnamefont
  {Guti{\'{e}}rrez-Lezama}}, \bibinfo {author} {\bibfnamefont {N.}~\bibnamefont
  {Ubrig}}, \bibinfo {author} {\bibfnamefont {M.}~\bibnamefont {Kroner}},
  \bibinfo {author} {\bibfnamefont {M.}~\bibnamefont {Gibertini}}, \bibinfo
  {author} {\bibfnamefont {T.}~\bibnamefont {Taniguchi}}, \bibinfo {author}
  {\bibfnamefont {K.}~\bibnamefont {Watanabe}}, \bibinfo {author}
  {\bibfnamefont {A.}~\bibnamefont {Imamo{\u{g}}lu}}, \bibinfo {author}
  {\bibfnamefont {E.}~\bibnamefont {Giannini}},\ and\ \bibinfo {author}
  {\bibfnamefont {A.~F.}\ \bibnamefont {Morpurgo}},\ }\bibfield  {title}
  {\bibinfo {title} {Very large tunneling magnetoresistance in layered magnetic
  semiconductor {CrI$_3$}},\ }\href
  {https://doi.org/10.1038/s41467-018-04953-8} {\bibfield  {journal} {\bibinfo
  {journal} {Nature Communications}\ }\textbf {\bibinfo {volume} {9}},\
  \bibinfo {pages} {2516} (\bibinfo {year} {2018}{\natexlab{b}})}\BibitemShut
  {NoStop}%
\bibitem [{\citenamefont {Brec}\ \emph {et~al.}(1985)\citenamefont {Brec},
  \citenamefont {Ouvrard},\ and\ \citenamefont {Rouxel}}]{Brec1985}%
  \BibitemOpen
  \bibfield  {author} {\bibinfo {author} {\bibfnamefont {R.}~\bibnamefont
  {Brec}}, \bibinfo {author} {\bibfnamefont {G.}~\bibnamefont {Ouvrard}},\ and\
  \bibinfo {author} {\bibfnamefont {J.}~\bibnamefont {Rouxel}},\ }\bibfield
  {title} {\bibinfo {title} {Relationship between structure parameters and
  chemical properties in some {MPS$_3$} layered phases},\ }\href
  {https://doi.org/10.1016/0025-5408(85)90118-7} {\bibfield  {journal}
  {\bibinfo  {journal} {Materials Research Bulletin}\ }\textbf {\bibinfo
  {volume} {20}},\ \bibinfo {pages} {1257} (\bibinfo {year}
  {1985})}\BibitemShut {NoStop}%
\bibitem [{\citenamefont {Susner}\ \emph {et~al.}(2017)\citenamefont {Susner},
  \citenamefont {Chyasnavichyus}, \citenamefont {McGuire}, \citenamefont
  {Ganesh},\ and\ \citenamefont {Maksymovych}}]{Susner2017}%
  \BibitemOpen
  \bibfield  {author} {\bibinfo {author} {\bibfnamefont {M.~A.}\ \bibnamefont
  {Susner}}, \bibinfo {author} {\bibfnamefont {M.}~\bibnamefont
  {Chyasnavichyus}}, \bibinfo {author} {\bibfnamefont {M.~A.}\ \bibnamefont
  {McGuire}}, \bibinfo {author} {\bibfnamefont {P.}~\bibnamefont {Ganesh}},\
  and\ \bibinfo {author} {\bibfnamefont {P.}~\bibnamefont {Maksymovych}},\
  }\bibfield  {title} {\bibinfo {title} {Metal thio- and selenophosphates as
  multifunctional van der waals layered materials},\ }\href
  {https://doi.org/10.1002/adma.201602852} {\bibfield  {journal} {\bibinfo
  {journal} {Advanced Materials}\ }\textbf {\bibinfo {volume} {29}},\ \bibinfo
  {pages} {1602852} (\bibinfo {year} {2017})}\BibitemShut {NoStop}%
\bibitem [{\citenamefont {Colombet}\ \emph {et~al.}(1982)\citenamefont
  {Colombet}, \citenamefont {Leblanc}, \citenamefont {Danot},\ and\
  \citenamefont {Rouxel}}]{PColombet1982}%
  \BibitemOpen
  \bibfield  {author} {\bibinfo {author} {\bibfnamefont {P.}~\bibnamefont
  {Colombet}}, \bibinfo {author} {\bibfnamefont {A.}~\bibnamefont {Leblanc}},
  \bibinfo {author} {\bibfnamefont {M.}~\bibnamefont {Danot}},\ and\ \bibinfo
  {author} {\bibfnamefont {J.}~\bibnamefont {Rouxel}},\ }\bibfield  {title}
  {\bibinfo {title} {Structural aspects and magnetic properties of the lamellar
  compound {Cu$_{0.50}$Cr$_{0.50}$PS$_{3}$}},\ }\href
  {https://doi.org/10.1016/0022-4596(82)90200-6} {\bibfield  {journal}
  {\bibinfo  {journal} {Journal of Solid State Chemistry}\ }\textbf {\bibinfo
  {volume} {41}},\ \bibinfo {pages} {174} (\bibinfo {year} {1982})}\BibitemShut
  {NoStop}%
\bibitem [{\citenamefont {Colombet}\ \emph {et~al.}(1983)\citenamefont
  {Colombet}, \citenamefont {Leblanc}, \citenamefont {Danot},\ and\
  \citenamefont {Rouxel}}]{PColombet1983}%
  \BibitemOpen
  \bibfield  {author} {\bibinfo {author} {\bibfnamefont {P.}~\bibnamefont
  {Colombet}}, \bibinfo {author} {\bibfnamefont {A.}~\bibnamefont {Leblanc}},
  \bibinfo {author} {\bibfnamefont {M.}~\bibnamefont {Danot}},\ and\ \bibinfo
  {author} {\bibfnamefont {J.}~\bibnamefont {Rouxel}},\ }\bibfield  {title}
  {\bibinfo {title} {Coordinance inhabituelle de l'argent dans un sufur
  lamellaire a sous-reseau magnetique {1D}: le compose
  {Ag$_{0.5}$Cr$_{0.5}$PS$_3$}},\ }\href@noop {} {\bibfield  {journal}
  {\bibinfo  {journal} {Nouveau Journal de Chimie}\ }\textbf {\bibinfo {volume}
  {7}},\ \bibinfo {pages} {333} (\bibinfo {year} {1983})}\BibitemShut {NoStop}%
\bibitem [{\citenamefont {Selter}\ \emph
  {et~al.}(2021{\natexlab{a}})\citenamefont {Selter}, \citenamefont
  {Shemerliuk}, \citenamefont {Büchner},\ and\ \citenamefont
  {Aswartham}}]{Selter2021a}%
  \BibitemOpen
  \bibfield  {author} {\bibinfo {author} {\bibfnamefont {S.}~\bibnamefont
  {Selter}}, \bibinfo {author} {\bibfnamefont {Y.}~\bibnamefont {Shemerliuk}},
  \bibinfo {author} {\bibfnamefont {B.}~\bibnamefont {Büchner}},\ and\
  \bibinfo {author} {\bibfnamefont {S.}~\bibnamefont {Aswartham}},\ }\bibfield
  {title} {\bibinfo {title} {Crystal growth of the quasi-{2D} quarternary
  compound {AgCrP$_2$S$_6$} by chemical vapor transport},\ }\href
  {https://doi.org/10.3390/cryst11050500} {\bibfield  {journal} {\bibinfo
  {journal} {Crystals}\ }\textbf {\bibinfo {volume} {11}},\ \bibinfo {pages}
  {500} (\bibinfo {year} {2021}{\natexlab{a}})}\BibitemShut {NoStop}%
\bibitem [{\citenamefont {Ouili}\ \emph {et~al.}(1987)\citenamefont {Ouili},
  \citenamefont {Leblanc},\ and\ \citenamefont {Colombet}}]{Ouili1987}%
  \BibitemOpen
  \bibfield  {author} {\bibinfo {author} {\bibfnamefont {Z.}~\bibnamefont
  {Ouili}}, \bibinfo {author} {\bibfnamefont {A.}~\bibnamefont {Leblanc}},\
  and\ \bibinfo {author} {\bibfnamefont {P.}~\bibnamefont {Colombet}},\
  }\bibfield  {title} {\bibinfo {title} {Crystal structure of a new lamellar
  compound: {Ag$_{12}$In$_{12}$PS$_3$}},\ }\href
  {https://doi.org/10.1016/0022-4596(87)90223-4} {\bibfield  {journal}
  {\bibinfo  {journal} {Journal of Solid State Chemistry}\ }\textbf {\bibinfo
  {volume} {66}},\ \bibinfo {pages} {86} (\bibinfo {year} {1987})}\BibitemShut
  {NoStop}%
\bibitem [{\citenamefont {Shannon}(1976)}]{RShannon1976}%
  \BibitemOpen
  \bibfield  {author} {\bibinfo {author} {\bibfnamefont {R.~D.}\ \bibnamefont
  {Shannon}},\ }\bibfield  {title} {\bibinfo {title} {Revised effective ionic
  radii and systematic studies of interatomic distances in halides and
  chalcogenides},\ }\href {https://doi.org/10.1107/s0567739476001551}
  {\bibfield  {journal} {\bibinfo  {journal} {Acta Crystallographica Section
  A}\ }\textbf {\bibinfo {volume} {32}},\ \bibinfo {pages} {751} (\bibinfo
  {year} {1976})}\BibitemShut {NoStop}%
\bibitem [{\citenamefont {Lai}\ \emph {et~al.}(2019)\citenamefont {Lai},
  \citenamefont {Song}, \citenamefont {Wan}, \citenamefont {Xue}, \citenamefont
  {Wang}, \citenamefont {Ye}, \citenamefont {Dai}, \citenamefont {Zhang},
  \citenamefont {Yang}, \citenamefont {Du},\ and\ \citenamefont
  {Yang}}]{Lai2019}%
  \BibitemOpen
  \bibfield  {author} {\bibinfo {author} {\bibfnamefont {Y.}~\bibnamefont
  {Lai}}, \bibinfo {author} {\bibfnamefont {Z.}~\bibnamefont {Song}}, \bibinfo
  {author} {\bibfnamefont {Y.}~\bibnamefont {Wan}}, \bibinfo {author}
  {\bibfnamefont {M.}~\bibnamefont {Xue}}, \bibinfo {author} {\bibfnamefont
  {C.}~\bibnamefont {Wang}}, \bibinfo {author} {\bibfnamefont {Y.}~\bibnamefont
  {Ye}}, \bibinfo {author} {\bibfnamefont {L.}~\bibnamefont {Dai}}, \bibinfo
  {author} {\bibfnamefont {Z.}~\bibnamefont {Zhang}}, \bibinfo {author}
  {\bibfnamefont {W.}~\bibnamefont {Yang}}, \bibinfo {author} {\bibfnamefont
  {H.}~\bibnamefont {Du}},\ and\ \bibinfo {author} {\bibfnamefont
  {J.}~\bibnamefont {Yang}},\ }\bibfield  {title} {\bibinfo {title}
  {Two-dimensional ferromagnetism and driven ferroelectricity in van der waals
  {CuCrP$_2$S$_6$}},\ }\href {https://doi.org/10.1039/c9nr00738e} {\bibfield
  {journal} {\bibinfo  {journal} {Nanoscale}\ }\textbf {\bibinfo {volume}
  {11}},\ \bibinfo {pages} {5163} (\bibinfo {year} {2019})}\BibitemShut
  {NoStop}%
\bibitem [{\citenamefont {Susner}\ \emph {et~al.}(2020)\citenamefont {Susner},
  \citenamefont {Rao}, \citenamefont {Pelton}, \citenamefont {McLeod},\ and\
  \citenamefont {Maruyama}}]{Susner2020}%
  \BibitemOpen
  \bibfield  {author} {\bibinfo {author} {\bibfnamefont {M.~A.}\ \bibnamefont
  {Susner}}, \bibinfo {author} {\bibfnamefont {R.}~\bibnamefont {Rao}},
  \bibinfo {author} {\bibfnamefont {A.~T.}\ \bibnamefont {Pelton}}, \bibinfo
  {author} {\bibfnamefont {M.~V.}\ \bibnamefont {McLeod}},\ and\ \bibinfo
  {author} {\bibfnamefont {B.}~\bibnamefont {Maruyama}},\ }\bibfield  {title}
  {\bibinfo {title} {Temperature-dependent {Raman} scattering and x-ray
  diffraction study of phase transitions in layered multiferroic
  {CuCrP$_2$S$_6$}},\ }\href
  {https://doi.org/10.1103/physrevmaterials.4.104003} {\bibfield  {journal}
  {\bibinfo  {journal} {Physical Review Materials}\ }\textbf {\bibinfo {volume}
  {4}},\ \bibinfo {pages} {104003} (\bibinfo {year} {2020})}\BibitemShut
  {NoStop}%
\bibitem [{\citenamefont {Kleemann}\ \emph {et~al.}(2011)\citenamefont
  {Kleemann}, \citenamefont {Shvartsman}, \citenamefont {Borisov},
  \citenamefont {Banys},\ and\ \citenamefont {Vysochanskii}}]{WKleemann2011}%
  \BibitemOpen
  \bibfield  {author} {\bibinfo {author} {\bibfnamefont {W.}~\bibnamefont
  {Kleemann}}, \bibinfo {author} {\bibfnamefont {V.~V.}\ \bibnamefont
  {Shvartsman}}, \bibinfo {author} {\bibfnamefont {P.}~\bibnamefont {Borisov}},
  \bibinfo {author} {\bibfnamefont {J.}~\bibnamefont {Banys}},\ and\ \bibinfo
  {author} {\bibfnamefont {Y.~M.}\ \bibnamefont {Vysochanskii}},\ }\bibfield
  {title} {\bibinfo {title} {Magnetic and polar phases and dynamical clustering
  in multiferroic layered solid solutions {CuCr$_{1-x}$In$_x$P$_2$S$_6$}},\
  }\href {https://doi.org/10.1103/physrevb.84.094411} {\bibfield  {journal}
  {\bibinfo  {journal} {Physical Review B}\ }\textbf {\bibinfo {volume} {84}},\
  \bibinfo {pages} {094411} (\bibinfo {year} {2011})}\BibitemShut {NoStop}%
\bibitem [{\citenamefont {Selter}\ \emph {et~al.}(2020)\citenamefont {Selter},
  \citenamefont {Bastien}, \citenamefont {Wolter}, \citenamefont {Aswartham},\
  and\ \citenamefont {Büchner}}]{SSelter2020}%
  \BibitemOpen
  \bibfield  {author} {\bibinfo {author} {\bibfnamefont {S.}~\bibnamefont
  {Selter}}, \bibinfo {author} {\bibfnamefont {G.}~\bibnamefont {Bastien}},
  \bibinfo {author} {\bibfnamefont {A.~U.~B.}\ \bibnamefont {Wolter}}, \bibinfo
  {author} {\bibfnamefont {S.}~\bibnamefont {Aswartham}},\ and\ \bibinfo
  {author} {\bibfnamefont {B.}~\bibnamefont {Büchner}},\ }\bibfield  {title}
  {\bibinfo {title} {Magnetic anisotropy and low-field magnetic phase diagram
  of the quasi-two-dimensional ferromagnet {Cr$_2$Ge$_2$Te$_6$}},\ }\href
  {https://doi.org/10.1103/physrevb.101.014440} {\bibfield  {journal} {\bibinfo
   {journal} {Physical Review B}\ }\textbf {\bibinfo {volume} {101}},\ \bibinfo
  {pages} {014440} (\bibinfo {year} {2020})}\BibitemShut {NoStop}%
\bibitem [{\citenamefont {Degen}\ \emph {et~al.}(2014)\citenamefont {Degen},
  \citenamefont {Sadki}, \citenamefont {Bron}, \citenamefont {König},\ and\
  \citenamefont {N{\'{e}}nert}}]{Degen2014}%
  \BibitemOpen
  \bibfield  {author} {\bibinfo {author} {\bibfnamefont {T.}~\bibnamefont
  {Degen}}, \bibinfo {author} {\bibfnamefont {M.}~\bibnamefont {Sadki}},
  \bibinfo {author} {\bibfnamefont {E.}~\bibnamefont {Bron}}, \bibinfo {author}
  {\bibfnamefont {U.}~\bibnamefont {König}},\ and\ \bibinfo {author}
  {\bibfnamefont {G.}~\bibnamefont {N{\'{e}}nert}},\ }\bibfield  {title}
  {\bibinfo {title} {The {HighScore} suite},\ }\href
  {https://doi.org/10.1017/s0885715614000840} {\bibfield  {journal} {\bibinfo
  {journal} {Powder Diffraction}\ }\textbf {\bibinfo {volume} {29}},\ \bibinfo
  {pages} {S13} (\bibinfo {year} {2014})}\BibitemShut {NoStop}%
\bibitem [{\citenamefont {Pet{\v{r}}{\'{\i}}{\v{c}}ek}\ \emph
  {et~al.}(2014)\citenamefont {Pet{\v{r}}{\'{\i}}{\v{c}}ek}, \citenamefont
  {Du{\v{s}}ek},\ and\ \citenamefont {Palatinus}}]{Petricek2014}%
  \BibitemOpen
  \bibfield  {author} {\bibinfo {author} {\bibfnamefont {V.}~\bibnamefont
  {Pet{\v{r}}{\'{\i}}{\v{c}}ek}}, \bibinfo {author} {\bibfnamefont
  {M.}~\bibnamefont {Du{\v{s}}ek}},\ and\ \bibinfo {author} {\bibfnamefont
  {L.}~\bibnamefont {Palatinus}},\ }\bibfield  {title} {\bibinfo {title}
  {Crystallographic computing system {JANA2006}: General features},\ }\href
  {https://doi.org/10.1515/zkri-2014-1737} {\bibfield  {journal} {\bibinfo
  {journal} {Zeitschrift für Kristallographie - Crystalline Materials}\
  }\textbf {\bibinfo {volume} {229}},\ \bibinfo {pages} {345} (\bibinfo {year}
  {2014})}\BibitemShut {NoStop}%
\bibitem [{\citenamefont {Helgaker}\ \emph {et~al.}(2000)\citenamefont
  {Helgaker}, \citenamefont {J{\o}rgensen},\ and\ \citenamefont
  {Olsen}}]{qc_book_00}%
  \BibitemOpen
  \bibfield  {author} {\bibinfo {author} {\bibfnamefont {T.}~\bibnamefont
  {Helgaker}}, \bibinfo {author} {\bibfnamefont {P.}~\bibnamefont
  {J{\o}rgensen}},\ and\ \bibinfo {author} {\bibfnamefont {J.}~\bibnamefont
  {Olsen}},\ }\href@noop {} {\emph {\bibinfo {title} {Molecular
  Electronic-Structure Theory}}}\ (\bibinfo  {publisher} {Wiley, Chichester},\
  \bibinfo {year} {2000})\BibitemShut {NoStop}%
\bibitem [{\citenamefont {Werner}\ \emph {et~al.}(2012)\citenamefont {Werner},
  \citenamefont {Knowles}, \citenamefont {Knizia}, \citenamefont {Manby},\ and\
  \citenamefont {Sch{\"u}tz}}]{Molpro}%
  \BibitemOpen
  \bibfield  {author} {\bibinfo {author} {\bibfnamefont {H.-J.}\ \bibnamefont
  {Werner}}, \bibinfo {author} {\bibfnamefont {P.~J.}\ \bibnamefont {Knowles}},
  \bibinfo {author} {\bibfnamefont {G.}~\bibnamefont {Knizia}}, \bibinfo
  {author} {\bibfnamefont {F.~R.}\ \bibnamefont {Manby}},\ and\ \bibinfo
  {author} {\bibfnamefont {M.}~\bibnamefont {Sch{\"u}tz}},\ }\bibfield  {title}
  {\bibinfo {title} {Molpro: a general-purpose quantum chemistry program
  package},\ }\href {https://doi.org/https://doi.org/10.1002/wcms.82}
  {\bibfield  {journal} {\bibinfo  {journal} {Wiley Rev.: Comp. Mol. Sci.}\
  }\textbf {\bibinfo {volume} {2}},\ \bibinfo {pages} {242} (\bibinfo {year}
  {2012})}\BibitemShut {NoStop}%
\bibitem [{\citenamefont {Selter}\ \emph
  {et~al.}(2021{\natexlab{b}})\citenamefont {Selter}, \citenamefont
  {Shemerliuk}, \citenamefont {Sturza}, \citenamefont {Wolter}, \citenamefont
  {Büchner},\ and\ \citenamefont {Aswartham}}]{Selter2021}%
  \BibitemOpen
  \bibfield  {author} {\bibinfo {author} {\bibfnamefont {S.}~\bibnamefont
  {Selter}}, \bibinfo {author} {\bibfnamefont {Y.}~\bibnamefont {Shemerliuk}},
  \bibinfo {author} {\bibfnamefont {M.-I.}\ \bibnamefont {Sturza}}, \bibinfo
  {author} {\bibfnamefont {A.~U.~B.}\ \bibnamefont {Wolter}}, \bibinfo {author}
  {\bibfnamefont {B.}~\bibnamefont {Büchner}},\ and\ \bibinfo {author}
  {\bibfnamefont {S.}~\bibnamefont {Aswartham}},\ }\bibfield  {title} {\bibinfo
  {title} {Crystal growth and anisotropic magnetic properties of
  quasi-two-dimensional {(Fe$_{1-x}$Ni$_x$)$_2$P$_2$S$_6$}},\ }\href
  {https://doi.org/10.1103/physrevmaterials.5.073401} {\bibfield  {journal}
  {\bibinfo  {journal} {Physical Review Materials}\ }\textbf {\bibinfo {volume}
  {5}},\ \bibinfo {pages} {073401} (\bibinfo {year}
  {2021}{\natexlab{b}})}\BibitemShut {NoStop}%
\bibitem [{\citenamefont {Dioguardi}\ \emph {et~al.}(2020)\citenamefont
  {Dioguardi}, \citenamefont {Selter}, \citenamefont {Peeck}, \citenamefont
  {Aswartham}, \citenamefont {Sturza}, \citenamefont {Murugesan}, \citenamefont
  {Eldeeb}, \citenamefont {Hozoi}, \citenamefont {Büchner},\ and\
  \citenamefont {Grafe}}]{Dioguardi2020}%
  \BibitemOpen
  \bibfield  {author} {\bibinfo {author} {\bibfnamefont {A.~P.}\ \bibnamefont
  {Dioguardi}}, \bibinfo {author} {\bibfnamefont {S.}~\bibnamefont {Selter}},
  \bibinfo {author} {\bibfnamefont {U.}~\bibnamefont {Peeck}}, \bibinfo
  {author} {\bibfnamefont {S.}~\bibnamefont {Aswartham}}, \bibinfo {author}
  {\bibfnamefont {M.-I.}\ \bibnamefont {Sturza}}, \bibinfo {author}
  {\bibfnamefont {R.}~\bibnamefont {Murugesan}}, \bibinfo {author}
  {\bibfnamefont {M.~S.}\ \bibnamefont {Eldeeb}}, \bibinfo {author}
  {\bibfnamefont {L.}~\bibnamefont {Hozoi}}, \bibinfo {author} {\bibfnamefont
  {B.}~\bibnamefont {Büchner}},\ and\ \bibinfo {author} {\bibfnamefont
  {H.-J.}\ \bibnamefont {Grafe}},\ }\bibfield  {title} {\bibinfo {title}
  {Quasi-two-dimensional magnetic correlations in {Ni$_2$P$_2$S$_6$} probed by
  $^{31}${P} {NMR}},\ }\href {https://doi.org/10.1103/physrevb.102.064429}
  {\bibfield  {journal} {\bibinfo  {journal} {Physical Review B}\ }\textbf
  {\bibinfo {volume} {102}},\ \bibinfo {pages} {064429} (\bibinfo {year}
  {2020})}\BibitemShut {NoStop}%
\bibitem [{\citenamefont {Brec}(1986)}]{RBrec1986}%
  \BibitemOpen
  \bibfield  {author} {\bibinfo {author} {\bibfnamefont {R.}~\bibnamefont
  {Brec}},\ }\bibfield  {title} {\bibinfo {title} {{R}eview on {S}tructural and
  {C}hemical {P}roperties of {T}ransition {M}etal {P}hosphorus {T}risulfides
  {MPS}$_{3}$},\ }in\ \href {https://doi.org/10.1007/978-1-4757-5556-5_4}
  {\emph {\bibinfo {booktitle} {{I}ntercalation in {L}ayered {M}aterials}}}\
  (\bibinfo  {publisher} {Springer {US}},\ \bibinfo {year} {1986})\ pp.\
  \bibinfo {pages} {93--124}\BibitemShut {NoStop}%
\bibitem [{\citenamefont {Shemerliuk}\ \emph {et~al.}(2021)\citenamefont
  {Shemerliuk}, \citenamefont {Zhou}, \citenamefont {Yang}, \citenamefont
  {Cao}, \citenamefont {Wolter}, \citenamefont {Büchner},\ and\ \citenamefont
  {Aswartham}}]{Shemerliuk2021}%
  \BibitemOpen
  \bibfield  {author} {\bibinfo {author} {\bibfnamefont {Y.}~\bibnamefont
  {Shemerliuk}}, \bibinfo {author} {\bibfnamefont {Y.}~\bibnamefont {Zhou}},
  \bibinfo {author} {\bibfnamefont {Z.}~\bibnamefont {Yang}}, \bibinfo {author}
  {\bibfnamefont {G.}~\bibnamefont {Cao}}, \bibinfo {author} {\bibfnamefont
  {A.~U.~B.}\ \bibnamefont {Wolter}}, \bibinfo {author} {\bibfnamefont
  {B.}~\bibnamefont {Büchner}},\ and\ \bibinfo {author} {\bibfnamefont
  {S.}~\bibnamefont {Aswartham}},\ }\bibfield  {title} {\bibinfo {title}
  {Tuning magnetic and transport properties in quasi-{2D}
  {(Mn$_{1-x}$Ni$_x$)$_2$P$_2$S$_6$} single crystals},\ }\href
  {https://doi.org/10.3390/electronicmat2030020} {\bibfield  {journal}
  {\bibinfo  {journal} {Electronic Materials}\ }\textbf {\bibinfo {volume}
  {2}},\ \bibinfo {pages} {284} (\bibinfo {year} {2021})}\BibitemShut {NoStop}%
\bibitem [{\citenamefont {Masubuchi}\ \emph {et~al.}(2008)\citenamefont
  {Masubuchi}, \citenamefont {Hoya}, \citenamefont {Watanabe}, \citenamefont
  {Takahashi}, \citenamefont {Ban}, \citenamefont {Ohkubo}, \citenamefont
  {Takase},\ and\ \citenamefont {Takano}}]{TMasubuchi2008}%
  \BibitemOpen
  \bibfield  {author} {\bibinfo {author} {\bibfnamefont {T.}~\bibnamefont
  {Masubuchi}}, \bibinfo {author} {\bibfnamefont {H.}~\bibnamefont {Hoya}},
  \bibinfo {author} {\bibfnamefont {T.}~\bibnamefont {Watanabe}}, \bibinfo
  {author} {\bibfnamefont {Y.}~\bibnamefont {Takahashi}}, \bibinfo {author}
  {\bibfnamefont {S.}~\bibnamefont {Ban}}, \bibinfo {author} {\bibfnamefont
  {N.}~\bibnamefont {Ohkubo}}, \bibinfo {author} {\bibfnamefont
  {K.}~\bibnamefont {Takase}},\ and\ \bibinfo {author} {\bibfnamefont
  {Y.}~\bibnamefont {Takano}},\ }\bibfield  {title} {\bibinfo {title} {Phase
  diagram, magnetic properties and specific heat of
  {Mn$_{1-x}$Fe$_{x}$PS$_{3}$}},\ }\href
  {https://doi.org/10.1016/j.jallcom.2007.06.063} {\bibfield  {journal}
  {\bibinfo  {journal} {Journal of Alloys and Compounds}\ }\textbf {\bibinfo
  {volume} {460}},\ \bibinfo {pages} {668} (\bibinfo {year}
  {2008})}\BibitemShut {NoStop}%
\bibitem [{\citenamefont {Klintenberg}\ \emph {et~al.}(2000)\citenamefont
  {Klintenberg}, \citenamefont {Derenzo},\ and\ \citenamefont
  {Weber}}]{Klintenberg_et_al}%
  \BibitemOpen
  \bibfield  {author} {\bibinfo {author} {\bibfnamefont {M.}~\bibnamefont
  {Klintenberg}}, \bibinfo {author} {\bibfnamefont {S.}~\bibnamefont
  {Derenzo}},\ and\ \bibinfo {author} {\bibfnamefont {M.}~\bibnamefont
  {Weber}},\ }\bibfield  {title} {\bibinfo {title} {Accurate crystal fields for
  embedded cluster calculations},\ }\href
  {https://doi.org/https://doi.org/10.1016/S0010-4655(00)00071-0} {\bibfield
  {journal} {\bibinfo  {journal} {Comp. Phys. Commun.}\ }\textbf {\bibinfo
  {volume} {131}},\ \bibinfo {pages} {120} (\bibinfo {year}
  {2000})}\BibitemShut {NoStop}%
\bibitem [{\citenamefont {Derenzo}\ \emph {et~al.}(2000)\citenamefont
  {Derenzo}, \citenamefont {Klintenberg},\ and\ \citenamefont
  {Weber}}]{Derenzo_et_al}%
  \BibitemOpen
  \bibfield  {author} {\bibinfo {author} {\bibfnamefont {S.~E.}\ \bibnamefont
  {Derenzo}}, \bibinfo {author} {\bibfnamefont {M.~K.}\ \bibnamefont
  {Klintenberg}},\ and\ \bibinfo {author} {\bibfnamefont {M.~J.}\ \bibnamefont
  {Weber}},\ }\bibfield  {title} {\bibinfo {title} {Determining point charge
  arrays that produce accurate ionic crystal fields for atomic cluster
  calculations},\ }\href {https://doi.org/10.1063/1.480776} {\bibfield
  {journal} {\bibinfo  {journal} {J. Chem. Phys.}\ }\textbf {\bibinfo {volume}
  {112}},\ \bibinfo {pages} {2074} (\bibinfo {year} {2000})}\BibitemShut
  {NoStop}%
\bibitem [{\citenamefont {Berning}\ \emph {et~al.}(2000)\citenamefont
  {Berning}, \citenamefont {Schweizer}, \citenamefont {Werner}, \citenamefont
  {Knowles},\ and\ \citenamefont {Palmieri}}]{SOC_molpro}%
  \BibitemOpen
  \bibfield  {author} {\bibinfo {author} {\bibfnamefont {A.}~\bibnamefont
  {Berning}}, \bibinfo {author} {\bibfnamefont {M.}~\bibnamefont {Schweizer}},
  \bibinfo {author} {\bibfnamefont {H.-J.}\ \bibnamefont {Werner}}, \bibinfo
  {author} {\bibfnamefont {P.~J.}\ \bibnamefont {Knowles}},\ and\ \bibinfo
  {author} {\bibfnamefont {P.}~\bibnamefont {Palmieri}},\ }\bibfield  {title}
  {\bibinfo {title} {Spin-orbit matrix elements for internally contracted
  multireference configuration interaction wavefunctions},\ }\href
  {https://doi.org/10.1080/00268970009483386} {\bibfield  {journal} {\bibinfo
  {journal} {Mol. Phys.}\ }\textbf {\bibinfo {volume} {98}},\ \bibinfo {pages}
  {1823} (\bibinfo {year} {2000})}\BibitemShut {NoStop}%
\bibitem [{\citenamefont {Balabanov}\ and\ \citenamefont
  {Peterson}(2005)}]{Balabanov_Cr}%
  \BibitemOpen
  \bibfield  {author} {\bibinfo {author} {\bibfnamefont {N.~B.}\ \bibnamefont
  {Balabanov}}\ and\ \bibinfo {author} {\bibfnamefont {K.~A.}\ \bibnamefont
  {Peterson}},\ }\bibfield  {title} {\bibinfo {title} {Systematically
  convergent basis sets for transition metals. i. all-electron correlation
  consistent basis sets for the 3d elements sc–zn},\ }\href
  {https://doi.org/10.1063/1.1998907} {\bibfield  {journal} {\bibinfo
  {journal} {J. Chem. Phys.}\ }\textbf {\bibinfo {volume} {123}},\ \bibinfo
  {pages} {064107} (\bibinfo {year} {2005})}\BibitemShut {NoStop}%
\bibitem [{\citenamefont {Woon}\ and\ \citenamefont
  {Dunning~Jr.}(1993)}]{Dunning_S}%
  \BibitemOpen
  \bibfield  {author} {\bibinfo {author} {\bibfnamefont {D.~E.}\ \bibnamefont
  {Woon}}\ and\ \bibinfo {author} {\bibfnamefont {T.~H.}\ \bibnamefont
  {Dunning~Jr.}},\ }\bibfield  {title} {\bibinfo {title} {Gaussian basis sets
  for use in correlated molecular calculations. iii. the atoms aluminum through
  argon},\ }\href {https://doi.org/10.1063/1.464303} {\bibfield  {journal}
  {\bibinfo  {journal} {J. Chem. Phys.}\ }\textbf {\bibinfo {volume} {98}},\
  \bibinfo {pages} {1358} (\bibinfo {year} {1993})}\BibitemShut {NoStop}%
\bibitem [{\citenamefont {Igel-Mann}\ \emph {et~al.}(1988)\citenamefont
  {Igel-Mann}, \citenamefont {Stoll},\ and\ \citenamefont
  {Preuss}}]{Igel-Mann_et_al}%
  \BibitemOpen
  \bibfield  {author} {\bibinfo {author} {\bibfnamefont {G.}~\bibnamefont
  {Igel-Mann}}, \bibinfo {author} {\bibfnamefont {H.}~\bibnamefont {Stoll}},\
  and\ \bibinfo {author} {\bibfnamefont {H.}~\bibnamefont {Preuss}},\
  }\bibfield  {title} {\bibinfo {title} {Pseudopotentials for main group
  elements (iiia through viia)},\ }\href
  {https://doi.org/10.1080/00268978800101811} {\bibfield  {journal} {\bibinfo
  {journal} {Mol. Phys.}\ }\textbf {\bibinfo {volume} {65}},\ \bibinfo {pages}
  {1321} (\bibinfo {year} {1988})}\BibitemShut {NoStop}%
\bibitem [{\citenamefont {Osborn}(1945)}]{JOsborn1945}%
  \BibitemOpen
  \bibfield  {author} {\bibinfo {author} {\bibfnamefont {J.~A.}\ \bibnamefont
  {Osborn}},\ }\bibfield  {title} {\bibinfo {title} {Demagnetizing factors of
  the general ellipsoid},\ }\href {https://doi.org/10.1103/physrev.67.351}
  {\bibfield  {journal} {\bibinfo  {journal} {Physical Review}\ }\textbf
  {\bibinfo {volume} {67}},\ \bibinfo {pages} {351} (\bibinfo {year}
  {1945})}\BibitemShut {NoStop}%
\bibitem [{\citenamefont {Moriya}\ \emph {et~al.}(2005)\citenamefont {Moriya},
  \citenamefont {Kariya}, \citenamefont {Inaba}, \citenamefont {Matsuo},
  \citenamefont {Pritz},\ and\ \citenamefont {Vysochanskii}}]{Moriya2005}%
  \BibitemOpen
  \bibfield  {author} {\bibinfo {author} {\bibfnamefont {K.}~\bibnamefont
  {Moriya}}, \bibinfo {author} {\bibfnamefont {N.}~\bibnamefont {Kariya}},
  \bibinfo {author} {\bibfnamefont {A.}~\bibnamefont {Inaba}}, \bibinfo
  {author} {\bibfnamefont {T.}~\bibnamefont {Matsuo}}, \bibinfo {author}
  {\bibfnamefont {I.}~\bibnamefont {Pritz}},\ and\ \bibinfo {author}
  {\bibfnamefont {Y.~M.}\ \bibnamefont {Vysochanskii}},\ }\bibfield  {title}
  {\bibinfo {title} {Low-temperature calorimetric study of phase transitions in
  {CuCrP$_2$S$_6$}},\ }\href
  {https://doi.org/https://doi.org/10.1016/j.ssc.2005.05.040} {\bibfield
  {journal} {\bibinfo  {journal} {Solid State Communications}\ }\textbf
  {\bibinfo {volume} {136}},\ \bibinfo {pages} {173} (\bibinfo {year}
  {2005})}\BibitemShut {NoStop}%
\bibitem [{\citenamefont {Cajipea}\ \emph {et~al.}(1996)\citenamefont
  {Cajipea}, \citenamefont {Ravez}, \citenamefont {Maisonneuve}, \citenamefont
  {Simon}, \citenamefont {Payen}, \citenamefont {Muhll},\ and\ \citenamefont
  {Fischer}}]{Cajipea1996}%
  \BibitemOpen
  \bibfield  {author} {\bibinfo {author} {\bibfnamefont {V.~B.}\ \bibnamefont
  {Cajipea}}, \bibinfo {author} {\bibfnamefont {J.}~\bibnamefont {Ravez}},
  \bibinfo {author} {\bibfnamefont {V.}~\bibnamefont {Maisonneuve}}, \bibinfo
  {author} {\bibfnamefont {A.}~\bibnamefont {Simon}}, \bibinfo {author}
  {\bibfnamefont {C.}~\bibnamefont {Payen}}, \bibinfo {author} {\bibfnamefont
  {R.~V.~D.}\ \bibnamefont {Muhll}},\ and\ \bibinfo {author} {\bibfnamefont
  {J.~E.}\ \bibnamefont {Fischer}},\ }\bibfield  {title} {\bibinfo {title}
  {Copper ordering in lamellar {CuMP$_2$S$_6$} ({M} = {Cr}, {In}): Transition
  to an antiferroelectric or ferroelectric phase},\ }\href
  {https://doi.org/10.1080/00150199608210497} {\bibfield  {journal} {\bibinfo
  {journal} {Ferroelectrics}\ }\textbf {\bibinfo {volume} {185}},\ \bibinfo
  {pages} {135} (\bibinfo {year} {1996})}\BibitemShut {NoStop}%
\bibitem [{\citenamefont {McGuire}\ \emph {et~al.}(2017)\citenamefont
  {McGuire}, \citenamefont {Clark}, \citenamefont {KC}, \citenamefont {Chance},
  \citenamefont {Jellison}, \citenamefont {Cooper}, \citenamefont {Xu},\ and\
  \citenamefont {Sales}}]{MMcGuire2017}%
  \BibitemOpen
  \bibfield  {author} {\bibinfo {author} {\bibfnamefont {M.~A.}\ \bibnamefont
  {McGuire}}, \bibinfo {author} {\bibfnamefont {G.}~\bibnamefont {Clark}},
  \bibinfo {author} {\bibfnamefont {S.}~\bibnamefont {KC}}, \bibinfo {author}
  {\bibfnamefont {W.~M.}\ \bibnamefont {Chance}}, \bibinfo {author}
  {\bibfnamefont {G.~E.}\ \bibnamefont {Jellison}}, \bibinfo {author}
  {\bibfnamefont {V.~R.}\ \bibnamefont {Cooper}}, \bibinfo {author}
  {\bibfnamefont {X.}~\bibnamefont {Xu}},\ and\ \bibinfo {author}
  {\bibfnamefont {B.~C.}\ \bibnamefont {Sales}},\ }\bibfield  {title} {\bibinfo
  {title} {Magnetic behavior and spin-lattice coupling in cleavable van der
  waals layered {CrCl$_3$} crystals},\ }\href
  {https://doi.org/10.1103/PhysRevMaterials.1.014001} {\bibfield  {journal}
  {\bibinfo  {journal} {Physical Review Materials}\ }\textbf {\bibinfo {volume}
  {1}},\ \bibinfo {pages} {014001} (\bibinfo {year} {2017})}\BibitemShut
  {NoStop}%
\bibitem [{\citenamefont {Zeisner}\ \emph {et~al.}(2020)\citenamefont
  {Zeisner}, \citenamefont {Mehlawat}, \citenamefont {Alfonsov}, \citenamefont
  {Roslova}, \citenamefont {Doert}, \citenamefont {Isaeva}, \citenamefont
  {Büchner},\ and\ \citenamefont {Kataev}}]{JZeisner2020}%
  \BibitemOpen
  \bibfield  {author} {\bibinfo {author} {\bibfnamefont {J.}~\bibnamefont
  {Zeisner}}, \bibinfo {author} {\bibfnamefont {K.}~\bibnamefont {Mehlawat}},
  \bibinfo {author} {\bibfnamefont {A.}~\bibnamefont {Alfonsov}}, \bibinfo
  {author} {\bibfnamefont {M.}~\bibnamefont {Roslova}}, \bibinfo {author}
  {\bibfnamefont {T.}~\bibnamefont {Doert}}, \bibinfo {author} {\bibfnamefont
  {A.}~\bibnamefont {Isaeva}}, \bibinfo {author} {\bibfnamefont
  {B.}~\bibnamefont {Büchner}},\ and\ \bibinfo {author} {\bibfnamefont
  {V.}~\bibnamefont {Kataev}},\ }\bibfield  {title} {\bibinfo {title} {Electron
  spin resonance and ferromagnetic resonance spectroscopy in the high-field
  phase of the van der waals magnet {CrCl$_3$}},\ }\href
  {https://doi.org/10.1103/physrevmaterials.4.064406} {\bibfield  {journal}
  {\bibinfo  {journal} {Physical Review Materials}\ }\textbf {\bibinfo {volume}
  {4}},\ \bibinfo {pages} {064406} (\bibinfo {year} {2020})}\BibitemShut
  {NoStop}%
\bibitem [{\citenamefont {Maisonneuve}\ \emph {et~al.}(1995)\citenamefont
  {Maisonneuve}, \citenamefont {Payen},\ and\ \citenamefont
  {Cajipe}}]{Maisonneuve1995}%
  \BibitemOpen
  \bibfield  {author} {\bibinfo {author} {\bibfnamefont {V.}~\bibnamefont
  {Maisonneuve}}, \bibinfo {author} {\bibfnamefont {C.}~\bibnamefont {Payen}},\
  and\ \bibinfo {author} {\bibfnamefont {V.}~\bibnamefont {Cajipe}},\
  }\bibfield  {title} {\bibinfo {title} {On {CuCrP$_2$S$_6$}: Copper disorder,
  stacking distortions, and magnetic ordering},\ }\href
  {https://doi.org/https://doi.org/10.1006/jssc.1995.1204} {\bibfield
  {journal} {\bibinfo  {journal} {Journal of Solid State Chemistry}\ }\textbf
  {\bibinfo {volume} {116}},\ \bibinfo {pages} {208} (\bibinfo {year}
  {1995})}\BibitemShut {NoStop}%
\end{thebibliography}%

\end{document}